\documentclass[11pt]{JHEP3new} 

\JHEPspecialurl{http://jhep.sissa.it/JOURNAL/JHEP3.tar.gz}

\usepackage{epsfig,multicol,bbm,rotating}

\newcommand\fverb{\setbox\pippobox=\hbox\bgroup\verb}
\newcommand\fverbdo{\egroup\medskip\noindent%
            \fbox{\unhbox\pippobox}\ }
\newcommand\fverbit{\egroup\item[\fbox{\unhbox\pippobox}]}
\newbox\pippobox

\def\ifm#1{\relax\ifmmode#1\else$#1$\fi}
\def\DAF{DA\char8NE}    
\def\x{\ifm{\times}}   \def\pt#1,#2,{\ifm{#1\x10^{#2}}}
\def\up#1{\ifm{^{#1}}}   \def\dn#1{\ifm{_{#1}}}  \def\plm{\ifm{\,\pm}\,}
\def\ab{\ifm{\sim}}   \def\deg{\ifm{^\circ}}
\def\gam{\ifm{\gamma}}  
\def\kl{\ifm{K_L}}   \def\ks{\ifm{K_S}}
\def\eiii{\ifm{\pi^\pm e^\mp\nu}}
\def\keiii{\ifm{K_{e3}}}
\def\muiii{\ifm{\pi^\pm \mu^\mp\nu}}
\def\kmuiii{\ifm{K_{\mu3}}}
\def\pio{\ifm{\pi^0\pi^0}}
\def\po{\ifm{\pi^0}}
\def\pic{\ifm{\pi^+\pi^-}}
\def\K{\ifm{K}}
\def\rmk{\rm\kern.5mm }
\def\f{\ifm{\phi}}
\def\eqq#1{(\ref{#1})}
\def\eq#1{eq.~(\ref{#1})}
   \def\minus{$-$}

\def\figb#1;#2;{\parbox{#2cm}{\epsfig{file=#1.eps,width=#2cm}}}

\newcommand{\bea}{\begin{eqnarray}}
\newcommand{\eea}{\end{eqnarray}}

\newcommand{\be}{\begin{equation}}
\newcommand{\ee}{\end{equation}}
\newcommand{\beq}{\begin{equation}}
\newcommand{\eeq}{\end{equation}}
\newcommand{\ba}{\begin{array}}
\newcommand{\ea}{\end{array}}
\newcommand{\beqa}{\begin{eqnarray}}
\newcommand{\eeqa}{\end{eqnarray}}
\newcommand{\no}{\nonumber}
\newcommand{\nn}{\nonumber}

\newcommand{\cO}{\mathcal{O}}
\newcommand{\cA}{\mathcal{A}}

\def\Re{{\rm Re}}
\def\Im{{\rm Im}}
\newcommand{\g}{\gamma}

\newcommand{\Ref}[1]{Ref.~\cite{#1}}
\newcommand{\Refs}[1]{Refs.~\cite{#1}}
\newcommand{\andRef}[1]{and~\cite{#1}}
\newcommand{\Fig}[1]{Fig.~\ref{#1}}
\newcommand{\Sec}[1]{Sec.~\ref{#1}}
\newcommand{\Tab}[1]{Table~\ref{#1}}
\newcommand{\Eq}[1]{Eq.~(\ref{#1})}
\newcommand{\etal}{et al.}

\newcommand{\BR}[1]{\ensuremath{{\rm BR}(#1)}}

\newcommand{\SN}[2]{\ensuremath{#1\times10^{#2}}}
\renewcommand{\vec}[1]{\ensuremath{\mathbf{#1}}}

\newcommand{\Vud}{\ensuremath{|V_{ud}|}}
\newcommand{\Vus}{\ensuremath{|V_{us}|}}
\newcommand{\Vusf}{\ensuremath{|V_{us}|f_+(0)}}
\newcommand{\Reps}{\ensuremath{{\rm Re}\:\epsilon'/\epsilon}}

\newcommand{\fp}{f_+(0)}
\newcommand{\gev}{{\rm GeV}}
\newcommand{\mev}{{\rm MeV}}
\newcommand{\text}{\rm}

\catcode`@=11 
\newdimen\z@ \z@=0pt 
\newskip\z@skip \z@skip=0pt plus0pt minus0pt
\def\m@th{\mathsurround=\z@}
\def\ialign{\everycr{}\tabskip\z@skip\halign} 
\def\eqalign#1{\null\,\vcenter{\openup\jot\m@th
  \ialign{\strut\hfil$\displaystyle{##}$&$\displaystyle{{}##}$\hfil
      \crcr#1\crcr}}\,}
\catcode`@=12 

\title{Precision tests of the Standard Model  with leptonic and semileptonic
        kaon decays}
\author{The FlaviaNet Kaon Working 
Group\footnote{WWW access at {\tt http://www.lnf.infn.it/wg/vus}}
\footnote{The members of the FlaviaNet Kaon Working
Group who contributed more significantly to this note are: 
M.~Antonelli, V.~Cirigliano,
P.~Franzini, S~Glazov, R.~Hill, G.~Isidori, F.~Mescia, M.~Moulson, M.~Palutan,
E.~Passemar, M.~Piccini, M.~Veltri, O.~Yushchenko, R.~Wanke.}
\footnote{ The Collaborations each take
responsibility for the preliminary results of their own experiment.}}

 \received{\today}      
\revised{\today}
\accepted{\today}       

\abstract{
We present a global analysis of leptonic and semileptonic
 kaon decays data, including all recent results by  BNL-E865,
 KLOE, KTeV, ISTRA+,  and NA48. Experimental results are
 critically reviewed and combined, taking into account theoretical
 (both analytical and numerical) constraints on the semileptonic
 kaon form factors. This analysis leads to a very
 accurate determination of $V_{us}$ and allows us to perform
 several stringent tests of the Standard Model. 
}


\begin{document}

\section{Introduction}
In the Standard Model, SM, transition rates of semileptonic processes such as 
$d^i \to  u^j \ell \nu$, with $d^i$ ($u^j$) being a generic 
down (up) quark, can be computed with high accuracy in terms 
of the Fermi coupling $G_F$ and the elements $V_{ji}$ of the 
Cabibbo-Kobayashi Maskawa (CKM) matrix~\cite{CKM}. 
Measurements of the transition rates provide therefore
 precise determinations of the fundamental SM couplings. 

A detailed analysis of semileptonic decays offers also 
the possibility to set stringent constraints on  new physics scenarios.
While within the SM all $d^i \to  u^j \ell \nu$ transitions 
are ruled by the same CKM coupling $V_{ji}$ (satisfying 
the unitarity condition $\sum_k |V_{ik}|^2 =1$) and 
$G_F$ is the same coupling appearing in the muon decay, 
this is not necessarily true beyond the SM. 
Setting bounds on the violations of CKM unitarity, 
violations of lepton universality, and deviations from 
the $V-A$ structure, allows us to put significant 
constraints on various new-physics scenarios
(or eventually find evidences of new physics). 

In the case of leptonic and semileptonic $K$ decays these tests 
are particularly significant given the large amount of
data recently collected by several experiments: 
BNL-E865, KLOE, KTeV, ISTRA+, and NA48. 
These data allow to perform very stringent SM tests which 
are almost free from hadronic uncertainties (such as the 
$\mu/e$ universality ratio in $K_{\ell 2}$ decays).
In addition, the high statistical precision  
and the detailed information on kinematical distributions  
have stimulated a substantial progress also on the theory side:
most of the theory-dominated errors associated to hadronic form 
factors have recently been reduced below the $1\%$ level. 

An illustration of the importance
of semileptonic $K$ decays  in testing the SM 
is provided by the unitarity relation 
\begin{equation}
|V_{ud}|^2 + |V_{us}|^2 +|V_{ub}|^2 = 1 + \epsilon_{\rm NP}~.
\label{eq:unitarity}
\end{equation}
Here the $V_{ji}$ are the CKM elements determined from 
the various $d^i \to  u^j$ processes, 
having fixed $G_F$ from the 
muon life time: $G_\mu = 1.166371 (6) \times 10^{-5} {\rm GeV}^{-2}$~\cite{mulan}.
 $\epsilon_{\rm NP}$ parametrizes possible 
deviations from the SM induced by dimension-six operators, 
contributing either to the muon decay or to the $d^i \to  u^j$ 
transitions. By dimensional arguments 
we expect $\epsilon_{\rm NP} \sim M^2_{W}/\Lambda_{\rm NP}^2$, 
where $\Lambda_{\rm NP}$ is the effective scale of new physics. 
The present accuracy on $|V_{us}|$, which is the dominant 
source of error in (\ref{eq:unitarity}), allows 
to set bounds on $\epsilon_{\rm NP}$ around  $0.1\%$
or equivalently to set bounds on the new physics scale well above 1~TeV.

In this note we report on  progress in the verification of the
relation~(\ref{eq:unitarity}) as well as on many other tests of the SM which 
can be performed with leptonic and semileptonic $K$ decays. 
The note is organized as follows. The phenomenological
framework needed to describe  $K_{\ell3}$ and $K_{\mu2}$ decays 
within and beyond the SM is briefly reviewed 
in Section~\ref{sec:Hsu}.  
Section\ref{sec:data} is dedicated to the combination of the experimental data.
The results and the interpretation are presented in 
Section~\ref{sec:results}.

\section{Theoretical framework}
\label{sec:Hsu}

\subsection{$K_{\ell3}$ and $K_{\ell2}$ rates within the SM}
\label{sect:KlSM}

Within the SM the photon-inclusive $K_{\ell3}$ and $K_{\ell2}$ decay
rates are conveniently decomposed as~\cite{Blucher:2005dc}
\bea
\label{eq:Mkl3}
\Gamma(K_{\ell 3(\gamma)}) &=&
{ G_F^2 m_K^5 \over 192 \pi^3} C_K
  S_{\rm ew}\,|V_{us}|^2 f_+(0)^2\,
I_K^\ell(\lambda_{+,0})\,\left(1 + \delta^{K}_{SU(2)}+\delta^{K \ell}_{\rm
em}\right)^2\,\quad ,\\
\frac{\Gamma(K^{\pm}_{\ell 2(\gamma)})}{\Gamma(\pi^{\pm}_{\ell 2(\gamma)})} &=&
\left|\frac{V_{us}}{V_{ud}} \right|^2\frac{f^2_K m_K}
{f^2_\pi m_\pi}\left(\frac{1-m^2_\ell/m_K^2}{1-m^2_\ell/m_\pi^2}\right)^2
\times\left(1+\delta_{\rm
em}\right)\quad ,\label{eq:Mkl2}
\eea
where $C_{K}=1$ ($1/2$) for the neutral (charged) kaon decays,
$I_K^\ell(\lambda_{+,0})$ is the phase space integral that  
depends on the (experimentally accessible) slopes of the 
form factors (generically denoted by  $\lambda_{+,\,0}$),
and $S_{\rm ew}=1.0232(3)$ is the universal short-distance 
electromagnetic correction computed in Ref.~\cite{Sirlin:1981ie}.
The channel-dependent long-distance electromagnetic 
correction factors are denoted by  $\delta_{\rm em}$ and  $\delta^{K \ell}_{\rm em}$.
In the $K_{\ell2}$ case  $\delta_{\rm em}=-0.0070(35)$~\cite{marcianokl2,ciriglianokl2}, 
while the four  $\delta^{K \ell}_{\rm em}$  are given in Table~\ref{tab:iso-brk},
together with the isospin-breaking corrections due to $m_u \not= m_d$, 
denoted by $\delta^{K}_{SU(2)}$.

The overall normalization of the $K_{\ell3}$ rates depends upon 
 $f_{+}(0)$,  the $K \to \pi$ vector form factor at  zero momentum transfer 
[$t=(p_K-p_\pi)^2 = 0$]. By convention,  $f_{+}(0)$ is defined 
for the $K^0 \to \pi^-$ matrix element, in the limit 
$m_u = m_d$ and $\alpha_{\rm em} \to 0$ (keeping kaon and pion masses to their physical value).
Similarly, $f_K/f_\pi$ is the ratio of the kaon and pion decay
constants defined in the $m_u = m_d$ and $\alpha_{\rm em} \to 0$ limit. 
The values of these hadronic parameters, which represent the 
dominant source of theoretical uncertainty, will be discussed in 
Sect.~\ref{sec:ffzero}.

\begin{table}[ht]
\setlength{\tabcolsep}{3.8pt}
\centering
\begin{tabular}{c||c||c|}
& $\delta^K_{SU(2)} (\%)$
& $\delta^{K \ell}_{\rm em}(\%) $   \\
\hline
$K^{0}_{e 3}$   &  0        &  +0.57(15)\\
$K^{+}_{e3}$    & 2.36(22)  &  +0.08(15)\\
$K^{0}_{\mu 3}$ &  0        &  +0.80(15)\\
$K^{+}_{\mu 3}$ & 2.36(22)  &  +0.05(15)
\end{tabular}
\caption{Summary of the isospin-breaking
corrections factors~\cite{Cirigliano,neufeld}. The electromagnetic
corrections factors correspond to the fully-inclusive 
$K_{\ell3(\gamma)}$ rate. }
\label{tab:iso-brk}
\end{table}

The errors for the $K_{\ell3}$ electromagnetic corrections, given in  
 Table~\ref{tab:iso-brk}, have been obtained
within ChPT, estimating higher-order corrections by naive dimensional 
analysis~\cite{Cirigliano,neufeld}. Higher-order chiral corrections 
have a minor impact in the breaking of lepton universality. The
 errors are correlated as given below:
\be
\left(
\begin{array}{cccc}
  1.0  &  0.1  &  0.8   & -0.1 \\
      &   1.0  & -0.1   &  0.8 \\
      &        &    1.0 &  0.1 \\
      &        &        &  1.0
\end{array}
\right)~.
\ee

\subsection{Parametrization of  $K_{\ell 3 }$ form factors}
\label{sec:ffpara}
The hadronic  $K \to \pi$ matrix element of the vector current
is described by two form factors (FFs),  $f_+(t)$ and  $f_0(t)$, defined by 
\be
\eqalign{
\langle\pi^{-}\left(  k\right)  |\bar{s}\gamma^{\mu}u|K^{0}\left(  p\right)
\rangle&= (p+k) ^\mu f_+^{}(t) +(p-k) ^\mu f_-^{}(t) \cr
f_-^{}(t)&=\frac{m^2_K-m^2_\pi}{t}\left(f_0^{}(t)  -f_+^{}(t)\right)\cr}
\label{Eq4}
\ee
where $t=(p-k)^2$. By construction,  $f_0(0)=f_+(0)$.

In order to compute the phase space integrals appearing in Eq.~(\ref{eq:Mkl3})
we need experimental or theoretical inputs about the  $t$-dependence
of $f_{+,0}(t)$. In principle,  Chiral Perturbation Theory (ChPT) 
and Lattice QCD are useful tools to set theoretical constraints.
However, in practice the  $t$-dependence of the FFs at present 
is better determined by measurements and by combining measurements 
and dispersion relations.

In the physical region, $\left( m_\ell^2<t<(m_K-m_\pi)^2 \right)$,
a very good approximation for the FFs is given by a Taylor expansion
 up to $t^2$ terms
\begin{equation}
  \tilde{f}_{+,0}(t) \equiv\frac{{f}_{+,\,0}(t)}{{f}_{+}(0)} = 1 + 
\lambda'_{+,0}~\frac{t}{m_\pi^2}+\frac{1}{2}\;\lambda''_{+,0}\,\left(\frac{t}{m_\pi^2}\right)^2\,
 + \ \dots .
  \label{eq:Taylor}
\end{equation}
Note that $t=(p_K-p_\pi)^2=m_K^2+m_\pi^2-2m_KE_\pi$, therefore the FFs depend only on $E_\pi$.
The FF parameters can thus be obtained from a fit to the pion spectrum which is of the form
$g(E_\pi) \times \tilde f(E_\pi)^2$. 
Unfortunately $t$ is maximum for $E_\pi$ = 0, where $g(E_\pi)$ vanishes.

Still, experimental information about the vector form factor $\tilde{f}_{+}$ measured both from 
$K_{e3}$ and $K_{\mu3}$ data are quite accurate and so far superior to theoretical predictions.
A pole parametrization, 
$\tilde{f}_{+}(t) = M_V^2/(M_V^2-t)$,
with $M_V \sim 892$ MeV 
corresponding to the $K^*(892)$ resonance and which predicts $\lambda''_{+}= 2(\lambda'_{+})^2$,
is in good agreement with present data (see later). Improvements of this 
parametrization have been proposed in Refs.~\cite{Hill, Moussallam:2007, bops08}.
 For instance, in Ref.~\cite{bops08}, a dispersive 
 parametrization for $\tilde{f}_+$, which has good analytical and unitarity
 properties and a correct threshold behavior, has been built.
 
 The situation for the scalar form factor $\tilde{f}_{0}(t)$ is more complex.
 For kinematical reasons $f_0(t)$ is only accessible from  $K_{\mu3}$ data
 and one has to deal with the correlations between the two form factors.
 Moreover, for  $f_0(t)$, 
 the curvature $\lambda''_{0}$ cannot be determined from the data and different
 assumptions for the parametrization of $\tilde {f}_{0}$ such as linear,
 quadratic or polar lead to different results for the slope $\lambda'_0$ which
 cannot be discriminated from the data alone. 
In turn, these ambiguities induce a systematic uncertainty for $V_{us}$,
 even though data for partial rates by itself are very accurate. 
 For this reason, the parametrization used has to rely on theoretical
 arguments being as model-independent as possible and allowing to measure
 at least the slope and the curvature of the form factor.

\subsubsection{Dispersive constraints}

The vector and scalar form factors $f_{+,0}(t)$ in Eq.~(\ref{Eq4})
are analytic functions in the complex
$t$--plane, except for a cut along the positive real axis, starting at the
first physical threshold $t_{\rm th} = (m_K+m_\pi)^2$, 
where they develop discontinuities. They are real for $t<t_{\rm th}$.

Cauchy's theorem implies that $f_{+,0}(t)$ can be written as 
a dispersive integral along the physical cut
\begin{equation}
\label{dis}
f_{+,0}(t) \; = \; \frac{1}{\pi} \int\limits^\infty_{t_{\rm th}}\!\! ds'\,
\frac{\Im f_{+,0}(s')}{(s'-t-i0)} + {\rm subtractions} \,,
\end{equation}
where all possible on-shell 
intermediate states contribute to its imaginary part $\Im F_k(s')$.
A number of subtractions is needed to make the integral convergent.
Particularly appealing is an improved dispersion 
relation recently proposed in Ref.~\cite{stern} 
where two subtractions are performed at $t=0$
(where by definition, $\tilde f_0(0)\equiv 1$) and at
the so-called Callan-Treiman point $t_{CT} \equiv (m_K^2-m_\pi^2)$ leading to 
\bea
\label{eq:Dispf}
\tilde f_0(t)&=&exp\left[\frac{t}{t_{CT}}
\left(\mathrm{ln}\left(\tilde f_0(t_{CT})\right)- G(t) \right) \right]\\ \nonumber
\mathrm{with}~~
G(t)&=&\frac{t_{CT}(t_{CT}-t)}{\pi}\int^\infty_{t_{\rm th}}\frac{ds'}{s'}
\frac{\phi(s')}{\left(s'-t_{CT}\right)\left(s'-t-i\epsilon\right)},
\eea
assuming that $\tilde f_0(t)$ has no zero.
Here $\phi(x)$, the phase of $\tilde f_0(t)$, can be identified in the elastic region 
with the S-wave, $I=1/2$ $K\pi$ scattering phase, 
$\delta_{K\pi}(s)$, according to Watson theorem. 

A subtraction at $t_{CT}$ has been performed 
because the Callan-Treiman theorem implies 
\be
\tilde f_0(t_{CT})=\frac{f_K}{f_\pi}\frac{1}{f_+(0)}+ \Delta_{CT},
\label{eq:CTrel}
\ee
where $\Delta_{CT} \sim  {\cal{O}} (m_{u,d}/4 \pi F_{\pi})$ is a small
 quantity.  ChPT estimates at NLO in the isospin
 limit~\cite{Gasser:1984}, obtain 
\be
\label{eq:DeltaCT}
\Delta_{CT}=(-3.5\pm 8)\times 10^{-3}~,
\ee
where the error is a conservative estimate of the high-order corrections to
the expansion in light quark masses~\cite{Leutwyler}. 
A complete  two-loop evaluation 
of $\Delta_{CT}$, consistent with this estimate,  
has been recently presented in Ref.~\cite{tal2}. 

Hence, with only one parameter, $\tilde f_0(t_{CT})$, 
one can determine the shape of $\tilde f_0$ 
by fitting the $K_{\mu3}$ decay 
distribution with the dispersive representation of $\tilde f_0(t)$, Eq.~(\ref{eq:Dispf}). 
Then, we can deduce from Eq.~(\ref{eq:Dispf}) 
the three first coefficients of the Taylor 
expansion, Eq.~(\ref{eq:Taylor}), see Ref.~\cite{stern}: 
\be
\label{eq:lbda'0}
\lambda'_0=\frac{m_\pi^2}{\Delta_{K\pi}}\left[ \mathrm{ln}\left(\tilde{f}_0(t_{CT})\right)-G(0) \right] =
\frac{m_\pi^2}{\Delta_{K\pi}}\left[\mathrm{ln}\left(\tilde{f}_0(t_{CT})\right)-0.0398(40))\right],
\ee
\be
\label{eq:lbda''0}
\lambda''_0= (\lambda'_0)^2  -2~ m^4_\pi/t_{CT}~G'(0) = (\lambda'_0)^2+ (4.16 \pm 0.50)\times 10^{-4}~,
\ee
\bea
\label{eq:lbda'''0}
\lambda'''_0 &=& (\lambda'_0)^3 -6~m^4_\pi/t_{CT}~G'(0)~\lambda'_0 -3 m_\pi^6/t_{CT}~G''(0) \nonumber \\
&=& (\lambda'_0)^3 + 3~(4.16 \pm 0.50) \times 10^{-4}~\lambda'_0+ (2.72 \pm 0.11) \times 10^{-5}.
\eea  
Furthermore, thanks to Eq.~(\ref{eq:CTrel}),
measuring $\tilde f_0(t_{CT})$ provides a significant 
constraint on  $f_K/f_\pi/f_+(0)$ limited only by the 
 small theoretical uncertainty on $\Delta_{CT}$.
As we will discuss in Section~\ref{sec:CTtest}, 
this represents a powerful consistency check of present 
lattice QCD estimates of $f_K/f_\pi$ and $f_+(0)$.

\medskip 

A similar dispersive parametrization for the vector form factor has been proposed 
in Ref.~\cite{bops08} with two subtractions performed at $t=0$. This leads to:
\be
\tilde f_+(t)=\exp\Bigl{[}\frac{t}{m_\pi^2}\left(\Lambda_+ + H(t)\right)\Bigr{]}~,~\mathrm{where}~~ 
H(t)=\frac{m_\pi^2t}{\pi} \int_{t_{K\pi}}^{\infty}
\frac{ds}{s^2}
\frac{\varphi (s)}
{(s-t-i\epsilon)}~.
\label{Dispfp}
\ee 
In the elastic region, the phase of the vector form factor, $\varphi(s)$, equals the $I=1/2$, P-wave $K\pi$ scattering phase.



\medskip

Additional tests can be performed using the expression for the scalar form factor $f_0(t)$ 
at order $p^6$ in ChPT~\cite{Bijnens:2003}:
\begin{equation}
\label{F0t}
f_0(t) \;=\; f_+(0) + \overline{\Delta}(t) + \frac{(f_K/f_\pi-1)}{m_K^2-m_\pi^2}\,t +
\frac{8}{f_\pi^4}\,(2C_{12}^r + C_{34}^r)\,(m_K^2+m_\pi^2) t -
\frac{8}{f_\pi^4}\,C_{12}^r\,t^2 \,,
\end{equation}
where
\bea
f_+(0)&=& 1+\Delta(0)-\,\frac{8}{f_\pi^4}\,(C_{12}^r+C_{34}^r)(m_K^2-m_\pi^2)^2\\ \nonumber
\lambda'_0&=&8\frac{m^2_\pi\left(m_\pi^2+m^2_K\right)}{f^4_\pi~f_+(0)}\left(2
C_{12}^r+C_{34}^r\right)
+ \frac{m^2_\pi}{m_K^2-m^2_\pi} \left(\frac{f_K}{f_\pi}\frac{1}{f_+(0)}
-\frac{1}{f_+(0)}\right) +m_\pi^2~\frac{\overline {\Delta}'(0)}{f_+(0)}\\ \nonumber
\lambda''_0&=&-16\frac{m^4_\pi}{f^4_\pi~f_+(0)}C_{12}^r+ m_\pi^4 \frac{\overline{\Delta}''(0)}{f_+(0)}
\eea
 Here $\overline\Delta(t)$ is a function
 which receives contributions from order $p^4$ and $p^6$, but like $\Delta(0)$
 it is independent of the $C_i^r$, and the order $p^4$ chiral constants $L_i^r$
 only appear at order $p^6$. $\overline\Delta(t)$ and $\Delta(0)$ have been evaluated in the
 physical region in Ref.~\cite{Bijnens:2003} 
 using for the $L_i^r$ values a fit to
 experimental data. 
An analysis has been presented in ref.~\cite{passe}.
However, the fit has to be reconsidered in light of the new experimental
results  as for instance considering the new $K_{\ell4}$ analysis from NA48 and
the updated value of $f_K/f_\pi$.

\subsubsection{Analyticity and improved series expansion}
\label{sect:z-theory}

Armed only with the knowledge that the form factor is analytic 
outside the cut on the real axis, analyticity provides powerful 
constraints on the form factor shape without recourse to model
assumptions.    
In particular, by an appropriate conformal mapping, 
the series expansion (2.5) necessarily ``resums'' into the 
form 
\begin{equation}\label{eq:aexpand}
f(t) = {1\over \phi} ( a_0 + a_1 z + a_2 z^2 + \dots ) \,,
\end{equation}
where $\phi$ is an analytic function and 
\begin{equation}
z(t,t_0) = {\sqrt{t_{th} - t} - \sqrt{t_{th} - t_0} \over
 \sqrt{t_{th} - t} + \sqrt{t_{th} - t_0} } 
\end{equation}
is the new expansion parameter. 
In this ``$z$ expansion'',
the factor $z(t,t_0)$ sums an infinite number of terms, transforming 
the original series, naively an expansion involving $t/t_+ \lesssim
0.3$, into a series with a much smaller expansion parameter.  For
example, the choice $t_0=t_{th}(1-\sqrt{1-(m_K^2-m_\pi^2)/t_{th}})$ 
minimizes the
maximum value of $z$ occurring in the physical region, and for
this choice $|z(t,t_0)| \lesssim 0.047$.  

The function $\phi$ and the
number $t_0$ may be regarded as defining a ``scheme'' for the
expansion.  The expansion parameter $z$ and coefficients $a_k$ are
then ``scheme-dependent'' quantities, with the scheme dependence
dropping out in physical observables such as $f(t)$.
For the vector form factor, a convenient choice for $\phi$
is 
\begin{eqnarray}
\label{eq:phiplus}
&&\phi_{F_+}(t,t_0,Q^2) = \sqrt{1\over 32\pi} {z(t,0)\over -t}
\left(z(t,-Q^2)\over -Q^2-t\right)^{3/2} \no \\
&& \times
\left(z(t,t_0)\over t_0-t\right)^{-1/2} 
\left(z(t,t_-)\over t_--t\right)^{-3/4} 
{t_+-t\over (t_+-t_0)^{1/4}} \,.
\end{eqnarray}
This choice is motivated by arguments of unitarity, 
whereby the coefficients can be bounded by calculating 
an inclusive production rate in perturbation theory~\cite{unitarity}.  
In fact, a much more stringent bound is obtained by isolating the
exclusive $K\pi$ production rate in the vector channel from 
$\tau$ decay data~\cite{Barate:1999hj}.  This enforces~\cite{Hill:2006bq} 
\begin{equation}\label{eq:bound13}
\sum_{k=0}^\infty {a_k^2 \over a_0^2} 
\lesssim 170 \,. 
\end{equation}
With this choice of $\phi$, and $Q^2=2\,{\rm GeV}^2$, 
a convenient choice for $t_0$ is $t_0=0.39\, (m_K-m_\pi)^2$.
This choice eliminates correlations in shape parameters $a_1/a_0$ 
and $a_2/a_0$. 
 
The bound on the expansion coefficients can be used to bound errors
on physical quantities describing the form factor shape, as discussed 
below in Sect.~\ref{sect:explsl}.
A similar expansion can be used for the scalar form factor. 
Note that error estimates based on 
(\ref{eq:bound13}) are conservative---no single coefficient is likely
to saturate the bound.  Also, this bound is a maximum taken over different 
schemes; more stringent bounds for particular schemes can be
found in \cite{Hill:2006bq}.   

In addition to the direct applications in $K_{\ell 3}$ decays, it 
is important for other purposes  
to constrain the first few coefficients in (\ref{eq:aexpand}), and
check whether the series converges as expected.   $K_{\ell 3}$ decays 
provide a unique opportunity to do this. 
For example, the same parameterization can be used to constrain the form factor 
shape in lattice calculations of $f(0)$, with the threshold $t_{th}$ adjusted
to the appropriate value for the simulated quark masses. 
Measurements of $a_k$ in the kaon system can similarly be used to  
confirm scaling arguments that apply also in the charm and bottom 
systems~\cite{Becher:2005bg}.

\subsection{$K_{\ell3}$ and $K_{\ell2}$ decays beyond the SM}
\label{sect:NP}

\subsubsection{The  $s\rightarrow u$ effective Hamiltonian}

On general grounds, assuming only Lorentz invariance and neglecting 
effective operators of dimension higher than six, $\Delta S=1$ 
charged-current transitions are described by  
10 independent operators:
\bea
H^{\Delta S=1}_{su}&=& -\frac{G_F}{\sqrt{2}} V_{us}
\left[c^{V}_{LL} (\bar s \gamma^\mu L u)(\bar \nu \g^\mu L \ell)
    +c^{V}_{LR} (\bar s \g^\mu L u)(\bar \nu \g^\mu R \ell)\right.\nn\\
&&\phantom{\frac{G_F\alpha}{4\pi}}
    +c^{V}_{RL} (\bar s \g^\mu R u)(\bar \nu \g^\mu L \ell)
    +c^{V}_{RR} (\bar s \g^\mu R u)(\bar \nu \g^\mu R \ell)\nn\\
&&\phantom{\frac{G_F\alpha}{4\pi}}
    +c^{S}_{LL} (\bar s L u)(\bar \nu L \ell)
    +c^{S}_{LR} (\bar s L u)(\bar \nu R \ell)\nn\\
&&\phantom{\frac{G_F\alpha}{4\pi}}
    +c^{S}_{RL} (\bar s R u)(\bar \nu L \ell)
    +c^{S}_{RR} (\bar s R u)(\bar \nu R \ell)\nn\\
&&\phantom{\frac{G_F\alpha}{4\pi}}
   \left. +c^{T}_{LL} (\bar s \sigma^{\mu\nu} L u)(\bar \nu \sigma^{\mu\nu} L \ell)
    +c^{T}_{RR} (\bar s \sigma^{\mu\nu} R u)(\bar \nu \sigma^{\mu\nu} R \ell)
\right] + {\rm h.c.}
\label{eq:ham1}
\eea
where $L=(1-\g_5)$ and $R=(1+\g_5)$. Defining this Hamiltonian at the 
weak scale, the SM case corresponds to $c^{V}_{LL}(M_W^2)=1$ and all 
the other coefficients set to zero.  The universal electromagnetic correction
factor $S_{\rm ew}$ appearing in Eq.~(\ref{eq:Mkl3})
describes the evolution of  $c^{V}_{LL}$ to hadronic scales:
$c^{V}_{LL}(M_\rho^2)/c^{V}_{LL}(M_W^2) = 1+(S_{\rm ew}-1)/2 \approx S_{\rm ew}^{1/2}$.
A similar expression can also be written for the Hamiltonian regulating 
$u\to d$  transitions.

In the case of $K\to \pi\,\ell\,\nu$ 
decays only six independent combinations 
of these operators have a non-vanishing tree-level matrix element:
\be
\eqalign{
\cA(K\to  \pi \ell \nu) &= \frac{G_F}{\sqrt{2}}V_{us}\,
\left\langle\pi\ell\nu\left|\vphantom{\frac{m_\ell}{M_W}}
 c_{V} (\bar s \g^\mu u)(\bar \nu \g_\mu \ell)+c_{A} (\bar s
\g^\mu u)(\bar \nu \g_\mu \g_5 \ell)\right.\right.\cr
&\kern0.2cm\left.\left. + \frac{m_\ell}{M_W} c_{S}(\bar s u)(\bar \nu \ell)
      + i\frac{m_\ell}{M_W} c_{P}(\bar s u)(\bar \nu \g_5 \ell)\right.\right.\cr
&\kern0.2cm\left.\left.+\frac{m_s m_\ell}{M_W^2}c^{T} (\bar s \sigma^{\mu\nu}  u)(\bar \nu \sigma^{\mu\nu} \ell)
    +\frac{m_s m_\ell}{M_W^2} c^{T}_{\g_5} (\bar s \sigma^{\mu\nu}  u)(\bar \nu \sigma^{\mu\nu}\g_5 \ell) + {\rm h.c.}\:
\right|K\right\rangle\cr}
\label{eq:ham}
\ee
where
\bea
c_{V}&=&+(c^{V}_{LL}+c^{V}_{RL}+c^{V}_{LR}+c^{V}_{RR})~,\\
c_{A}&=&-(c^{V}_{LL}+c^{V}_{RL}-c^{V}_{LR}-c^{V}_{RR})~,\\
c_{S}&=&+(c^{S}_{LL}+c^{S}_{RL}+c^{S}_{LR}+c^{S}_{RR})M_W/m_\ell~,\\
i c_{P}&=&-(c^{S}_{LL}+c^{S}_{RL}-c^{S}_{LR}-c^{S}_{RR})M_W/m_\ell~,\\
c^{T}&=&2(c^{T}_{LL}+c^{T}_{RR})M^2_W/(m_\ell m_s)~,\quad
c^{T}_{\g_5}=-2(c^{T}_{LL}-c^{T}_{RR})M^2_W/(m_\ell m_s)~.
\eea
Similarly, in the $K\to \ell \nu$ case the independent structures are
\be\eqalign{
\cA(K\to \ell \nu)
&=-\frac{G_F}{\sqrt{2}}V_{us}\,
\left\langle\ell\nu\left|\vphantom{\frac{m_\ell}{M_W}}k_{V} (\bar s \g^\mu\g_5 u)(\bar \nu \g_\mu \ell)+k_{A} (\bar s
\g^\mu\g_5 u)(\bar \nu \g_\mu\g_5  \ell)\right.\right.\cr
&\left.\left. + \frac{m_\ell}{M_W} k_{S}(\bar s \g_5 u)(\bar \nu\ell)
+ \frac{m_\ell}{M_W} k_{P}(\bar s \g_5 u)(\bar \nu \g_5 \ell) + {\rm h.c.} \:
\right|K\right\rangle\cr}
\label{eq:hamkl2}
\ee
where
\bea
k_{V}&=&-(c^{V}_{LL}-c^{V}_{RL}+c^{V}_{LR}-c^{V}_{RR})~,\\
k_{A}&=&+(c^{V}_{LL}-c^{V}_{RL}-c^{V}_{LR}+c^{V}_{RR})~,\\
k_{S}&=&-(c^{S}_{LL}-c^{S}_{RL}+c^{S}_{LR}-c^{S}_{RR})M_W/m_\ell~,\\
k_{P}&=&+(c^{S}_{LL}-c^{S}_{RL}-c^{S}_{LR}+c^{S}_{RR})M_W/m_\ell~.
\eea

On general grounds, new degrees of freedom weakly coupled 
 at the scale $\Lambda_{\rm NP}$ are 
expected to generate corrections of $\cO(M^2_{W}/\Lambda^2_{\rm NP})$ 
to the Wilson coefficients of $H^{\Delta S=1}_{su}$.
Focusing on well-motivated new-physics frameworks, the 
following two scenarios are particularly interesting:
\begin{itemize}
\item
In two Higgs doublet models of type-II, such as the Higgs sector of the MSSM, 
sizable contributions are potentially generated by charged-Higgs exchange diagrams 
(see e.g.~Ref.~\cite{gino,retico,paride}). 
These are well described by the following set of initial conditions 
for $s\to u$ transitions,
\be
c^{V}_{LL}=1\,\qquad{\rm{and}}\qquad
c^{S}_{LR}=-\frac{\tan^2\beta}{\left(1+\epsilon_0\tan\beta\right)}
\frac{m_{\ell}m_s}{m^2_{H^+}}\, ,
\ee
and for $u\to d$ transitions,
\be
c^{V,ud}_{LL}=1\,\qquad{\rm{and}}\qquad
c^{S,ud}_{LR}=-\frac{\tan^2\beta}{\left(1+\epsilon_0\tan\beta\right)}
\frac{m_{\ell} m_d}{m^2_{H^+}}\,.
\ee
Here $\tan\beta$ is the ratio of the two Higgs vacuum expectation values 
and $\epsilon_0$ is a loop 
function whose detailed expression 
can be found in Ref.~\cite{retico}. In presence of sizable sources 
of lepton-flavor symmetry breaking, a non-vanishing scalar-current 
contribution to the lepton-flavor violating 
process $K\to e \nu_\tau$ is also present~\cite{paride}. The latter
can be parametrized by 
\be
c^{S^\prime}_{LR}=\frac{m_s m_\tau}{m^2_{H^+}}\Delta_R^{31}\tan^2\beta~.
\ee

\item
In the Higgs-less model of Ref.~\cite{stern}, non-standard 
right-handed quark currents could become detectable. These 
are described by the following set of initial conditions
 for both 
$u\to s$ and $u\to d$ transitions
\be
c^{V}_{LL}=\left(1+\delta\right)\,\qquad{\rm{and}}\qquad
c^{V}_{RL}=\,\epsilon_s\, ,
\ee
\be
c^{V,ud}_{LL}=\left(1+\delta\right)\,\qquad{\rm{and}}\qquad
c^{V,ud}_{RL}=\,\epsilon_{ns}\, ,
\ee
where $\varepsilon_x$ and $\delta$ are 
free parameters of the model. 
$\epsilon_s$ can reach a few percents if
the hierarchy of the right-handed mixing matrix is inverted.
\end{itemize}

\subsubsection{$K_{\ell 2}$ rates}
According to the Hamiltonian of  Eq.~(\ref{eq:hamkl2}), 
the $K_{\ell 2}$ rate of Eq.~(\ref{eq:Mkl2}) can be
modified as
\be\eqalign{
\frac{\Gamma(K^{\pm}_{\ell 2(\gamma)})}{\Gamma(\pi^{\pm}_{\ell 2(\gamma)})} =
\left|\frac{V_{us}}{V_{ud}}\right|^2& \frac{f^2_K m_K}
{f^2_\pi m_\pi}\left(\frac{1-m^2_\ell/m_K^2}{1-m^2_\ell/m_\pi^2}\right)^2
\times\left(1+\delta_{\rm
em}\right)\cr
&\kern0.2cm\times \frac{|k_A-m_K^2/(m_s M_W)k_P|^2 + |k_V+m_K^2/(m_s M_W)k_S|^2}
{|k^{ud}_A-m_\pi^2/(\hat m M_W)k^{ud}_P|^2 + |k^{ud}_V+m_\pi^2/(\hat m
M_W)k^{ud}_S|^2}\,,
\cr}\label{eq:Mkl2all}
\ee
where $\hat m= m_u+m_d$ and $k^{ud}_x$ are defined for the $u\to d$ transition. 
In the MSSM scenario
\beq
\Gamma^{\rm MSSM}(K_{\ell 2})/\Gamma^{\rm MSSM}(\pi_{\ell 2})= 
\Gamma^{\rm SM}(K_{\ell 2})/\Gamma^{\rm SM}(\pi_{\ell 2}) \times (1-r^K_H)^2~,
\label{eq:kl2mssm}
\eeq
where 
\beq
r^K_H= \frac{m^2_{K^+}}{M^2_{H^+}}\left(1 - \frac{m_d}{m_s}\right)
\frac{\tan^2\beta}{1+\epsilon_0\tan\beta}
\label{eq:rk}
\eeq

\subsubsection{$K_{\ell 3}$ rates and kinematical distributions}

In the  $K_{\ell 3}$ case the non-standard operators of Eq.~(\ref{eq:ham})
could in principle modify the Daliz plot distribution.
However, as we will show in the following, this effect 
turns out to be hardly detectable for most realistic new-physics 
scenarios. 

The hadronic form factors needed in the general case 
are the two FFs defined in Eq.~(\ref{Eq4}) plus a tensor FF, 
whereas $f_{0}(t)$ allow us to parametrize also the 
scalar-current  matrix element. More specifically, we  have
\bea
\langle\pi^-\left(k\right)|(\bar{s}u)|K^0\left(  p\right)\rangle&=&
-\frac{m_{K}^{2}-m_{\pi}^{2}}{\left(  m_{s}-m_{u}\right)  }f_{0}\left( 
t\right)~,\\
\langle\pi^-\left(k\right)|(\bar s \sigma^{\mu\nu} u)\vert K^0\left(p\right)
\rangle &=&i\,\frac{p^{\mu} k^{\nu} -p^{\nu}k^{\mu}}{m_{K}}
B_{T}\left(t\right)\,.
\label{SP1}
\eea
The tensor form-factor was studied on the lattice
\cite{Mescia00}, with the result $B_{T}(t)  \approx
1.2(1)\,f_{+}\left(  0\right)  /(1-0.3(1)t)$ 
at $\mu\simeq 2\gev$ in the
$\overline{MS}$ scheme (an earlier order-of-magnitude estimate may be found in
Ref.\cite{OldBT}). 

Choosing as independent kinematical variables 
$$ z= \frac{2 p_K \cdot p_\pi}{m_K^2}=\frac{m_K^2+m_\pi^2-t}{m_K^2},   \qquad  
y= \frac{2 p_K \cdot p_\ell }{m_K^2},  \qquad r_{\pi, \ell} = \frac{m_{\pi, \ell}^2}{m_K^2} $$
the double differential 
density can be written as (neglecting long-distance electromagnetic corrections)
\begin{eqnarray}
\frac{d \Gamma}{dy \,  dz}  &=& 
\frac{G_F^2 \vert V_{us}\vert^2  m_K^5}{256 \pi^3} C_K S_{\rm ew}
\, \Bigg[    A_1 (y,z)  \,   \left( |V|^2 + |A|^2 \right)  - 
A_2 (y,z) \,  {\rm Re} \left(V S^* - A P^* \right)  
\nonumber \\
&+& A_3 (y,z)   \, \left( |S|^2 + |P|^2 \right)      \Bigg]
\end{eqnarray} 
whereas 
\begin{eqnarray}
A_1 (y,z) &=& 4 (z + y - 1) (1 - y) 
+ r_\ell (4 y + 3 z - 3) - 4 r_\pi + r_\ell (r_\pi - r_\ell)    , \nonumber \\
A_2  (y,z) & = & 2 r_\ell (3 - 2 y - z + r_\ell - r_\pi)  , \qquad
A_3  (y,z) =   r_\ell ( 1 + r_\pi - z - r_\ell). 
\end{eqnarray}
Here $S$, $P$, $V$, and $A$ are convenient combinations of 
hadronic form factors and short-distance Wilson coefficients:
\bea
\label{eq:VASPT}
V(t,y) &=&     f_+(t)\,c_V -m_\ell^2 \frac{m_s}{M_W^2} \frac{c_{T} B_{T}(t)}{m_K}   \\
A(t,y)  &  =&  f_+(t) \,c_A + m_\ell^2 \frac{m_s}{M_W^2} \frac{c_{T5} B_{T}(t)}{m_K}  \no \\
S(t,y)  &  =& -( f^S_{0}(t)-f_+(t))\frac{m_K^2 - m_\pi^2}{t}\,c_V
-\left(m_\ell^2+m^2_K\left(2-z-2\,y\right)\right)
\frac{m_s}{M_W^2}\frac{c_{T} B_{T}(t)}{m_K} \no \\
P(t,y)  &  =& (f^P_{0}(t)- f_+(t))\frac{m_K^2 - m_\pi^2}{t}\,c_A
-\left(m_\ell^2+m^2_K\left(2-z-2\,y\right)\right)
\frac{m_s}{M_W^2}\frac{c_{T5} B_{T}(t)}{m_K},\no 
\eea
where 
\bea
f^{S}_0(t)&=&
f_0(t)\left(1
    +  \frac{c_S/c_V}{(m_s-m_u) M_W } \,t
    \right)\approx f_0(t)\,\exp\left(\frac{c_S/c_V}{M_W}\frac{m^2_K-m^2_\pi}
{m_s-m_u}\right)^{t/t_{CT}} \\
f^{P}_0(t)&=&
f_0(t)\left(1
    -  \frac{i\,c_P/c_A}{(m_s-m_u) M_W}\,t
    \right)\approx f_0(t)\,\exp\left(\frac{-i\,c_P/c_A}{M_W}
\frac{m^2_K-m^2_\pi}{m_s-m_u}\right)^{t/t_{CT}}, 
\eea
$t_{CT}=(m^2_K-m^2_\pi)$ and we have assumed $c_{S,P}/c_{V,A} \ll 1$.
The SM case is recovered from Eq.~(\ref{eq:VASPT})
in the limit $c_V=-c_A=1$ and $f^{S,P}_0(t)=f_0(t)$.

After integrating over $y$, differences to the SM rate of Eq.~(\ref{eq:Mkl3})
can be summarized as it follows. 
Right-handed currents can only rescale the overall
rate of Eq.~(\ref{eq:Mkl3}), namely
\begin{equation}
\Gamma(K_{\ell 3(\gamma)}) \to \Gamma(K_{\ell 3(\gamma)})\x 
\frac{\textstyle\vert c_V\vert^2+\vert c_A\vert^2}{\textstyle 2}\,.
\end{equation}
Scalar and pseudoscalar contributions
can be easily encoded in Eq.~(\ref{eq:Mkl3}) by  substituting
\begin{equation}
f_0(t) \to 
f^H_0(t) = f_0(t)\,\exp\left(\frac{\left(-i\,c_P c^*_A+c_S c^*_V\right)}{
\vert c_V\vert^2 +\vert c_A\vert^2}
\frac{m^2_K-m^2_\pi}{M_W\,m_s}\right)^{t/t_{CT}}~.
\end{equation}
In particular, these new effects are vanishing for $t=0$,  
namely $f_0(0)$ in Eq.~(\ref{eq:Mkl3}) is free from them. 
The tensor coupling modify the phase space integral 
$I_K^\ell(\lambda_{+,0})$ of Eq.~(\ref{eq:Mkl3}) by  
\begin{equation}
I_K^\ell(\lambda_{+,0}) \to I_K^\ell(\lambda_{+,0}) - 
\frac{\Re(c^T c_V^*)-\Re(c^T_{\g_5} c^*_A)}{\vert c_V\vert^2 +\vert c_A\vert^2} I_T^\ell(\lambda_{T,+,0})
\end{equation}
In conclusion, the integrated rate including  electromagnetic corrections can be 
written as
\bea
\label{eq:Mkl3all}
\Gamma(K_{\ell 3(\gamma)}) &=&
{ G_F^2 m_K^5 \over 192 \pi^3} C_K\,
  S_{\rm ew}\,|V_{us}|^2 f_+(0)^2\,\left(1 + \delta^{K}_{SU(2)}+\delta^{K \ell}_{\rm
em}\right)^2\\
&\x&\frac{\textstyle\vert c_V\vert^2+\vert c_A\vert^2}{\textstyle 2}
\left( {I}_K^\ell - 
\frac{\Re(c^T c_V^*)-
\Re(c^T_{\g_5} c^*_A)}{\vert c_V\vert^2 +\vert c_A\vert^2} I_T^\ell\right)\nn
\eea
where 
\bea
{ I}^\ell_K = \frac{\textstyle 1}{\textstyle m_K^2 f_+(0)^2} &&\!\int dt ~
\lambda^{3/2}(t) ~
  \left( 1+ \frac{\textstyle m_{\ell}^2}{\textstyle 2t} \right)
\left( 1 - \frac{\textstyle m_{\ell}^2} {\textstyle t} \right)^2\\ 
&&\times
 \left({f}^2_+(t) + \frac{\textstyle 3m_{\ell}^2
\left(m_K^2-m_\pi^2\right)^2}{\textstyle \left( 2t+m_{\ell}^2 \right)
m_K^4 \lambda(t)  }
\  \vert {f}^H_0(t)\vert^2
\right) \no\,,
\eea
\bea
{ I}^\ell_T = \frac{\textstyle 1}{\textstyle m_K^2 f_+(0)^2} &&\!\int dt ~ 
\lambda^{3/2}(t) ~ 
\frac{\textstyle m_\ell}{4 \textstyle m_K} 
\left( 1+ \frac{\textstyle 2 m_{\ell}^2}{\textstyle t} \right)
\left( 1 - \frac{\textstyle m_{\ell}^2} {\textstyle t} \right)^2 \\
&&B_T(t)
\times\left(
{f}_+(t)  + 
6 \frac{2 m_\ell^2(m_K^2 - m_\pi^2)^2
-(m_K^4 - m_\pi^4)t + t^2}{(t+2 m_\ell^2)m_K^4\lambda(t)}{f}_0(t)
\right)\nn
\eea
and $\lambda(t)=1 - 2 r_\pi^2 + r_\pi^4 - 2 t/m_K^2 - 2 r_\pi^2 t/m_K^2 
+ t^2/m_K^4$.

In most realistic new-physics scenarios the modification 
of the $K_{\ell 3}$ scalar form factor is well below 
the present experimental and theoretical errors. 
For instance, in the MSSM (or two-Higgs doublets) 
case $f^H_0(t)$ reads
\bea
\left. f^H_0(t)  \right|_{\rm MSSM} &=& f_0(t)\,
\exp\left(-r^K_H\right)^{t/t_{CT}}
\label{eq:f0mssm}
\eea
where $r^K_H$
is the parameter controlling the corrections 
to the $K_{\ell 2}$ rate of Eq.~(\ref{eq:kl2mssm}).
For natural values of the free parameters
($\epsilon_0=10^{-2}$, $M^2_{H^+}=400$~GeV and
$\tan\beta=40$), such that $r^K_H = 0.2\%$, 
the corresponding modification of the 
$K_{\ell 3}$ scalar form factor is 
\beq
\frac{\delta \lambda_{0}}{\lambda^{\rm SM}_{0}} \approx 1.0\% 
\qquad {\rm or} \qquad 
\frac{ \delta f_0(t_{CT})}{ f_0(t_{CT})^{\rm SM} } 
\approx 0.18 \%~, 
\eeq
well below the level of present 
theoretical and experimental uncertainties.

\section{Data Analysis}
\label{sec:data}
We perform fits to world data on the BRs and lifetimes for the
$K_L$ and $K^\pm$, with the constraint that BRs add to unity.
This is the correct way of using
 the new measurements. The fit procedure is described
 in Appendix~\ref{app:fitprocedure}.
\subsection{$K_L$   leading branching ratios and $\tau_L$ }
\label{sec:KL}
Numerous measurements of the principal $K_L$ BRs, or of various ratios
of these BRs, have been published recently. For the purposes of evaluating
\Vusf, these data can be used in a PDG-like fit to the $K_L$ BRs and lifetime,
so all such measurements are interesting. A detailed description to
the fit of the principal $K_L$ BRs  and $\tau_L$ is given in Appendix~\ref{app:BRLfit}

KTeV has measured five ratios of the six main $K_L$ BRs~\cite{KTeV+04:BR}.
The six channels
involved account for more than 99.9\% of the $K_L$ width and KTeV combines the
five measured ratios to extract the six BRs. We use the five measured ratios
in our analysis:
$\BR{K_{\mu3}}/\BR{K_{e3}} = 0.6640(26)$,
$\BR{\pi^+\pi^-\pi^0}/\BR{K_{e3}} = 0.3078(18)$,
$\BR{\pi^+\pi^-}/\BR{K_{e3}} = 0.004856(28)$,
$\BR{3\pi^0}/\BR{K_{e3}} = 0.4782(55)$, and
$\BR{2\pi^0}/\BR{3\pi^0} = 0.004446(25)$. The errors on these measurements are
correlated; this is taken into account in our fit.

NA48 has measured the ratio of the BR for $K_{e3}$ decays to the sum of BRs
for all decays to two tracks, giving
$\BR{K_{e3}}/(1-\BR{3\pi^0}) = 0.4978(35)$ \cite{NA48+04:BR}. From a
separate measurement of \BR{K_L\to3\pi^0}/\BR{K_S\to2\pi^0}, NA48
obtains $\BR{3\pi^0}/\tau_L = 3.795(58)$ $\mu$s\up{-1} \cite{Lit04:ICHEP}.

Using $\phi\to K_L K_S$ decays in which the $K_S$ decays to $\pi^+\pi^-$,
providing normalization, KLOE has directly measured the BRs for the four
main $K_L$ decay channels \cite{KLOE+06:BR}.
The errors on the KLOE BR values are dominated
by the uncertainty on the $K_L$ lifetime $\tau_L$; since the dependence of
the geometrical efficiency on $\tau_L$ is known, KLOE can solve for $\tau_L$
by imposing $\sum_x \BR{K_L\to x} = 1$ (using previous averages for the minor
BRs), thereby greatly reducing the uncertainties on the BR values obtained.
Our fit makes use of the KLOE BR values before application of this constraint:
\BR{K_{e3}} = 0.4049(21),
\BR{K_{\mu3}} = 0.2726(16),
\BR{K_{e3}} = 0.2018(24), and
\BR{K_{e3}} = 0.1276(15).
The dependence of these values on $\tau_L$ and the correlations between the
errors  are taken into account.
KLOE has also measured $\tau_L$ directly, by fitting the proper decay time
distribution for $K_L\to3\pi^0$ events, for which the reconstruction
efficiency is high and uniform over a fiducial volume of $\sim$$0.4\lambda_L$.
They obtain $\tau_L=50.92(30)$~ns \cite{KLOE+05:tauL}.

There are also two recent measurements of \BR{\pi^+\pi^-}/\BR{K_{\ell3}},
in addition to the KTeV measurement of \BR{\pi^+\pi^-}/\BR{K_{e3}} discussed above.
The KLOE collaboration 
obtains \BR{\pi^+\pi^-}/\BR{K_{\mu3}} = \SN{7.275(68)}{-3} \cite{KLOE+06:KLpp},
while NA48 obtains \BR{\pi^+\pi^-}/\BR{K_{e3}} = \SN{4.826(27)}{-3}
\cite{NA48+06:KLpp}. All measurements are fully inclusive of inner
bremsstrahlung. The KLOE measurement is fully inclusive of the direct-emission
(DE) component, DE contributes negligibly to the KTeV measurement, and a
residual DE contribution of 0.19\% has been subtracted from the NA48 value
to obtain the number quoted above. For consistency, in our fit,
a DE contribution of 1.52(7)\% is added to the KTeV and NA48 values.
Our fit result for \BR{\pi^+\pi^-} is then understood to be DE inclusive.

In addition to the 14 recent measurements listed above, our fit for the
seven largest $K_L$ BRs and lifetime uses four of the remaining five
inputs to the 2006 PDG fit and the constraint that the seven BRs add to unity.
The results are given in \Tab{tab:KLBR}.

\begin{table}
\begin{center}
\begin{tabular}{l|c|r}
Parameter & Value & $S$ \\
\hline
\BR{K_{e3}} & 0.4056(7) & 1.1 \\
\BR{K_{\mu3}} & 0.2705(7) & 1.1 \\
\BR{3\pi^0} & 0.1951(9) & 1.2 \\
\BR{\pi^+\pi^-\pi^0} & 0.1254(6) & 1.1 \\
\BR{\pi^+\pi^-} & \SN{1.997(7)}{-3} & 1.1 \\
\BR{2\pi^0} & \SN{8.64(4)}{-4} & 1.3 \\
\BR{\gamma\gamma} & \SN{5.47(4)}{-4} & 1.1 \\
$\tau_L$ & 51.17(20)~ns & 1.1 \\
\end{tabular}
\end{center}
\vskip 0.3cm
\caption{\label{tab:KLBR}
Results of fit to $K_L$ BRs and lifetime.}
\end{table}
The evolution of the average values of the BRs for
$K_{L\ell3}$ decays and for
the important normalization channels is shown in \Fig{fig:kpmavg}.

\begin{figure}[ht]
\begin{center}
\includegraphics[width=0.9\textwidth]{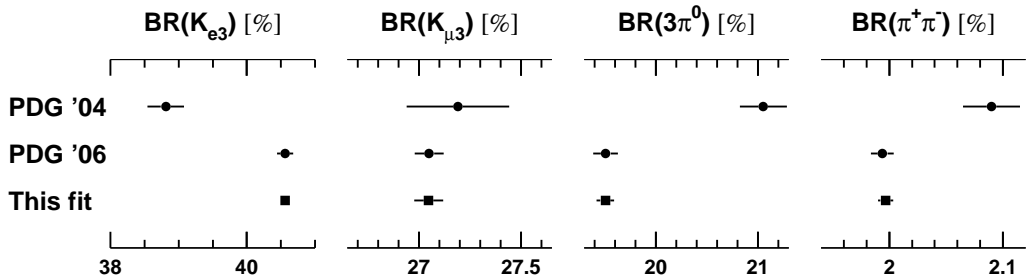}
\end{center}
\caption{\label{fig:klavg}
Evolution of average values for main $K_L$ BRs.}
\end{figure}

Our fit gives $\chi^2/{\rm ndf} = 20.2/11$ (4.3\%), while the 2006
PDG fit gives $\chi^2/{\rm ndf} = 14.8/10$ (14.0\%).
The differences between the output values from our fit and the
2006 PDG fit are minor.
The poorer value of $\chi^2/{\rm ndf}$ for our fit can be traced
to contrast between the KLOE value for \BR{3\pi^0} and the
other inputs involving \BR{3\pi^0} and \BR{\pi^0\pi^0}---in
particular, the PDG ETAFIT value for \BR{\pi^0\pi^0/\pi^+\pi^-}.
The treatment of the correlated KTeV and KLOE measurements in the
2006 PDG fit gives rise to large
scale factors for \BR{K_{e3}} and \BR{3\pi^0};
in our fit, the scale factors are more uniform. As a result,
our value for \BR{K_{e3}} has a significantly smaller uncertainty
than does the 2006 PDG value.

\subsection{$K_S$  leading branching ratios and $\tau_S$}
\label{sec:KS}
KLOE has published \cite{KLOE+06:KSe3}
a measurement of \BR{K_S\to\pi e\nu} that is precise enough
to contribute meaningfully to the evaluation of \Vusf.
The quantity directly measured is \BR{\pi e\nu}/\BR{\pi^+\pi^-}. Together
with the  published KLOE value
\BR{\pi^+\pi^-}/\BR{\pi^0\pi^0} = 2.2459(54),
the constraint that the $K_S$ BRs must add to unity, and the assumption of
universal lepton couplings, this completely determines the $K_S$ BRs for
$\pi^+\pi^-$, $\pi^0\pi^0$, $K_{e3}$, and $K_{\mu3}$ decays
\cite{KLOE+06:KSpp}. In particular, $\BR{K_S\to\pi e\nu} = \SN{7.046(91)}{-4}$.

NA48 has recently measured the ratio
$\Gamma(K_S \to \pi e \nu)/\Gamma(K_L \to \pi e \nu) = 0.993(26)(22)$
 \cite{NA48:KSe3}.
The best way to include this measurement in our analysis would
be via a combined fit to $K_S$ and $K_L$ branching ratio and lifetime
measurements. Indeed, such a fit would be useful in properly
accounting for correlations between $K_S$ and $K_L$ modes introduced
with the preliminary NA48 measurement of $\Gamma(K_L\to 3\pi^0)$, and
more importantly, via the PDG ETAFIT result, which we use in the
fit to $K_L$ branching ratios. At the moment, however, we fit
$K_S$ and $K_L$ data separately. NA48 quotes
$\BR{K_S\to\pi e\nu} = \SN{7.046(180)(160)}{-4}$;
averaging this with the KLOE result gives
$\BR{K_S\to\pi e\nu} = \SN{7.05(8)}{-4}$,
 improving the accuracy on this BR by about 10\%. 

For $\tau_{K_S}$ we use \SN{0.8958}{-10}~s, where this is the non-$CPT$
constrained fit value from the PDG, and is dominated by the 2002 NA48
and 2003 KTeV measurements.

\subsection{$K^\pm$ leading branching ratios and $\tau^\pm$}
There are several new results providing information on $K^\pm_{\ell3}$
rates. These results are mostly preliminary and have not been included
in previous averages.

The NA48/2 collaboration has recently  published measurements of the three ratios
\BR{K_{e3}/\pi\pi^0}, \BR{K_{\mu3}/\pi\pi^0}, and
\BR{K_{\mu3}/K_{e3}} \cite{NA48+07:BR}.
These measurements are not independent; in our fit, we use the values
$\BR{K_{e3}/\pi\pi^0} = 0.2470(10)$ and
$\BR{K_{\mu3}/\pi\pi^0} = 0.1637(7)$ and take their correlation
into account.
ISTRA+ has also updated its preliminary value for $\BR{K_{e3}/\pi\pi^0}$.
They now quote $\BR{K_{e3}/\pi\pi^0} = 0.2449(16)$\cite{Rom06:ke3}.

KLOE has measured the absolute BRs for the
$K_{e3}$ and $K_{\mu3}$ decays
\cite{KLOE:kl3pm}.
In $\phi\to K^+ K^-$ events, $K^+$ decays into $\mu\nu$ or $\pi\pi^0$
are used to tag a $K^-$ beam, and vice versa. KLOE performs four
separate measurements for each $K_{\ell3}$ BR, corresponding to the
different combinations of kaon charge and tagging decay.
The final averages are $\BR{K_{e3}} = 4.965(53)\%$ and
$\BR{K_{\mu3}} = 3.233(39)\%$.
Very recently KLOE has also measured the absolute
branching ratio for the $\pi\pi^0$ decay with 0.5\% accuracy.
The KLOE preliminary result, is  $\BR{\pi\pi^0}=0.20658(112)$\cite{KLOE:pipo}.

 Our fit takes into account the correlation between these values, as
 well as their dependence on the $K^\pm$ lifetime.
 The world average value for $\tau_\pm$ is nominally
 quite precise; the 2006 PDG quotes $\tau_\pm = 12.385(25)$~ns.
 However, the error is scaled by 2.1; the confidence level for the
 average is 0.17\%. It is important to confirm the value of $\tau_\pm$.
 The two new measurements from KLOE,
 $\tau_\pm = 12.367(44)(65)$~ns  and
 $\tau_\pm = 12.391(49)(25)$~ns\cite{KLOE:taupm} with correlation 34\%,
 agree with the PDG average.

 Our fit for the six largest $K^\pm$ BRs and lifetime makes use of the
 results cited above,
 plus the data used in the 2006 PDG fit, except for the
 Chiang '72 measurements
for a total of 26 measurements.
The six BRs are constrained to add to unity.
The results are shown in \Tab{tab:KpmBR}.
\begin{table}
\begin{center}
\begin{tabular}{l|c|r}
Parameter & Value & $S$ \\
\hline
\BR{K_{\mu2}}      & 63.57(11)\%   & 1.1 \\
\BR{\pi\pi^0}      & 20.64(8)\%   & 1.1 \\
\BR{\pi\pi\pi}     &  5.595(31)\%  & 1.0 \\
\BR{K_{e3}}        &  5.078(26)\%    & 1.2 \\
\BR{K_{\mu3}}      &  3.365(27)\%  & 1.7 \\
\BR{\pi\pi^0\pi^0} &  1.750(26)\%  & 1.1 \\
$\tau_\pm$         & 12.384(19)~ns & 1.7 \\
\end{tabular}
\end{center}
\vskip 0.3cm
\caption{\label{tab:KpmBR}
Results of fit to $K^\pm$ BRs and lifetime.}
\end{table}

\begin{figure}[ht]
\begin{center}
\includegraphics[width=0.9\textwidth]{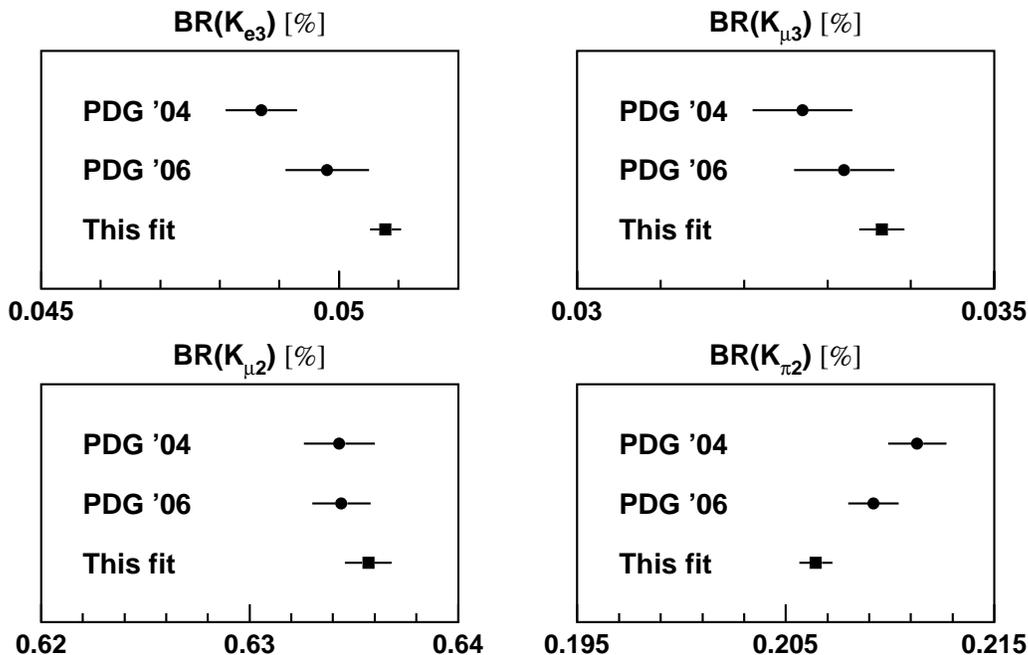}
\end{center}
\caption{\label{fig:kpmavg}
Evolution of average values for main $K^\pm$ BRs.}
\end{figure}

The fit quality is poor, with $\chi^2/{\rm ndf} = 42/20$ (0.31\%).
However, when the five older measurements of $\tau_\pm$ are replaced
by their PDG average with scaled error,  $\chi^2/{\rm ndf}$ improves
to 24.3/16 (8.4\%), with no significant changes in the results.

Both the significant evolution of the average values of the $K_{\ell3}$
BRs and the effect of the correlations with \BR{\pi\pi^0} are evident
in \Fig{fig:kpmavg}.

\subsection{Measurement of BR($K_{e2})$/BR($K_{\mu2}$)}
Experimental knowledge of $K_{e2}/K_{\mu2}$ has been poor so far.
 The current world average
of $R_K = \BR{K_{e2}}/\BR{K_{\mu2}}= (2.45 \pm 0.11) \times 10^{-5}$ dates back to three experiments
of the 1970s~\cite{bib:pdg} and has a precision of about 5\%.
Three new preliminary measurements were reported by NA48/2 and KLOE
 (see Tab.~\ref{tab:ke2kmu2}):
A preliminary result of NA48/2, based on about 4000 $K_{e2}$ events from the 2003
 data set, was presented in 2005~\cite{bib:Ke2_2003}.
Another preliminary result, based on also about 4000 events, recorded in a minimum bias
 run period in 2004, was shown at KAON07\cite{bib:Ke2_2004}.
Both results have independent statistics and are also independent in the systematic uncertainties,
as the systematics are either of statistical nature (as e.g.\ trigger efficiencies) or determined in
an independent way.
Another preliminary result, based on about 8000 $K_{e2}$ events, was presented at KAON07
by the KLOE collaboration~\cite{bib:Ke2_KLOE}.
Both, the KLOE and the NA48/2 measurements are inclusive with respect to final state radiation  contribution due to bremsstrahlung.
The small contribution of $K_{l2\gamma}$ events from direct photon emission from the decay vertex was subtracted by each of the experiments.
Combining these new results with the current PDG value yields a current world average of
\begin{equation}
R_K  = ( 2.457 \pm 0.032 ) \times 10^{-5},
\label{eqn:ke2kmu2}
\end{equation}
in very good agreement with the SM expectation and, with a relative error of $1.3\%$,
a factor three more precise than the previous world average.

\begin{table}[t]
  \begin{center}
      \begin{tabular}{lc}
        \hline \hline
                                                  & $R_K$ $[10^{-5}]$  \\ \hline
        PDG 2006~\cite{bib:pdg}                   & $2.45 \pm 0.11$ \\
        NA48/2 prel.\ ('03)~\cite{bib:Ke2_2003}   & $2.416 \pm 0.043 \pm 0.024$ \\
        NA48/2 prel.\ ('04)~\cite{bib:Ke2_2004}   & $2.455 \pm 0.045 \pm 0.041$ \\
        KLOE prel.~\cite{bib:Ke2_KLOE}            & $2.55 \pm 0.05 \pm 0.05$ \\ \hline
        SM prediction                             & $2.477 \pm 0.001$ \\
        \hline \hline
                                                  & \\*[-3mm]
      \end{tabular}
      \caption{Results and prediction for $R_K =  \BR{K_{e2}}/\BR{K_{\mu2}}$.}
      \label{tab:ke2kmu2}
  \end{center}
\end{table}

\subsection{\mathversion{bold}Measurements of $K_{\ell3}$ slopes}
\label{sect:explsl}

\subsubsection{\mathversion{bold}Vector form factor slopes from  $K_{\ell3}$}

For $K_{e3}$ decays, recent measurements of the quadratic slope parameters
of the vector form factor $({\lambda_+',\lambda_+''})$ are available from
KTeV \cite{KTeV+04:FF},
KLOE \cite{KLOE+06:FF}, ISTRA+ \cite{ISTRA+04:e3FF}, and
NA48 \cite{NA48+04:e3FF}.

We show the results of a fit to the $K_L$ and $K^-$ data in the
first column of \Tab{tab:e3ff}, and to only the $K_L$ data in the
second column. With correlations correctly
taken into account, both fits give good
values of $\chi^2/{\rm ndf}$. The significance of the quadratic
term is $4.2\sigma$ from the fit to all data, and $3.5\sigma$ from
the fit to $K_L$ data only.
\TABLE{
\begin{tabular}{lcc}
\hline\hline
& $K_L$ and $K^-$ data & $K_L$ data only \\
& 4 measurements & 3 measurements \\
& $\chi^2/{\rm ndf} = 5.3/6$ (51\%) &
  $\chi^2/{\rm ndf} = 4.7/4$ (32\%)\\
\hline
\SN{\lambda_+'}{3}
     & $25.2\pm0.9$ & $24.9\pm1.1$ \\
\SN{\lambda_+''}{3}
     & $1.6\pm0.4$ & $1.6\pm0.5$ \\
$\rho(\lambda_+',\lambda+'')$
     & $-0.94$ & $-0.95$ \\
$I(K^0_{e3})$
     & 0.15465(24) & 0.15456(31) \\
$I(K^\pm_{e3})$
     & 0.15901(24) & 0.15891(32) \\
\hline\hline
\end{tabular}
\caption{Average of quadratic fit results for $K_{e3}$ slopes.}
\label{tab:e3ff}
}

Including or excluding the $K^-$ slopes
has little impact on the values of $\lambda_+'$ and $\lambda_+''$;
in particular, the values of the phase-space integrals change by just
0.07\%. The errors on the phase-space integrals are significantly
smaller when the $K^-$ data are included in the average. 

KLOE, KTeV, and NA48 also quote the values shown in \Tab{tab:pole}
for $M_V$ from pole fits to $K_{L\:e3}$ data. The average value of
$M_V$ from all three experiments is
$M_V = 875\pm5$~MeV with $\chi^2/{\rm ndf} = 1.8/2$.
The three values are quite compatible with each other and
reasonably close to the known value of the $K^{\pm*}(892)$
mass ($891.66\pm0.26$~MeV). The values for $\lambda_+'$ and $\lambda_+''$
from expansion of the pole parametrization are qualitatively in
agreement with the average of the quadratic fit results.
More importantly, for the evaluation of the phase-space
integrals, using the average of quadratic or pole fit results gives
values of $I(K^0_{e3})$ that differ by just 0.03\%.
%
%
%
\TABLE{
\begin{tabular}{lc|c}
\hline\hline
Experiment & $M_V$ (MeV) & $\left<M_V\right> = 875\pm5$ MeV \\
KLOE & $870\pm6\pm7$ &  $\chi^2/{\rm ndf} = 1.8/2$ \\
KTeV & $881.03\pm7.11$ & \SN{\lambda_+'}{3} = 25.42(31) \\
NA48 & $859\pm18$ & $\lambda_+''=2\times\lambda_+'^{\,2}$ \\
& &  $I(K^0_{e3})$ = 0.15470(19) \\
\hline\hline
\end{tabular}
\caption{Pole fit results for $K^0_{e3}$ slopes.}
\label{tab:pole}}

An attempt to estimate the theoretical uncertainties associated to 
form factor parameterization has been pursued by KTeV,
analyzing  $K_{e3}$ decays with the $z$-expansion~\ref{sect:z-theory} 
for the $f_+(t)$ form factor~\cite{ktevz}.
The results are $a_1/a_0 = 1.023 \pm 0.040$ and  $a_2/a_0 = 0.75 \pm 2.16$.
The second order term is consistent with zero and the higher orders are bounded
by the theory:  $\sum_{k=0}^{\infty}a_k^2/a_o^2 \le 170$. 
Using these results the phase space integral is calculated to be 
$I(K^0_{e3})=0.15392 \pm 0.00048_{\rm exp} \pm 0.00006_{\rm th}$.
The first error corresponds to the KTeV experimental uncertainty and the second error
is due to  possible effects from higher order terms in the $z$-expansion.
Compared to the global average using the quadratic parameterization (Table 5),
the KTeV measurement using the $z$-expansion deviates by about $1.5\sigma_{\rm exp}$.
This result is less precise statistically, but it is more conservative
as far as the estimate of the theoretical uncertainty is concerned.

\subsubsection{Scalar and Vector form factor slopes from \boldmath{$K_{\ell3}$}}
\label{sec:l3ff}
 For $K_{\mu3}$ decays, recent measurements of the
 slope parameters $({\lambda_+',\lambda_+'',\lambda_0})$ are
 available from  KTeV \cite{KTeV+04:FF},  KLOE \cite{KLOE+07:m3FF},
 ISTRA+ \cite{ISTRA+04:m3FF}, and  NA48 \cite{NA48+06:m3FF}.
 These data are summarized in Appendix~\ref{app:ff}.

 We have studied the statistical sensitivity of the
 form-factor slope measurements using Monte Carlo techniques,
 see Appendix~\ref{app:Fslope}.. The conclusions of this study are a) 
 that neglecting a quadratic term in the parameterization of the 
 scalar form factor when fitting results  leads to  a shift of the value
 of the linear term by about 3.5 times the value of the quadratic
 term; and b)
 that because of correlations, it is impossible to measure the 
 quadratic slope parameter from quadratic fits to the data at 
 any plausible level of statistics.
 The use of the linear representation of the scalar form factor 
 is thus inherently unsatisfactory.

\begin{figure}[ht]
\begin{center}
\includegraphics[width=0.95\textwidth]{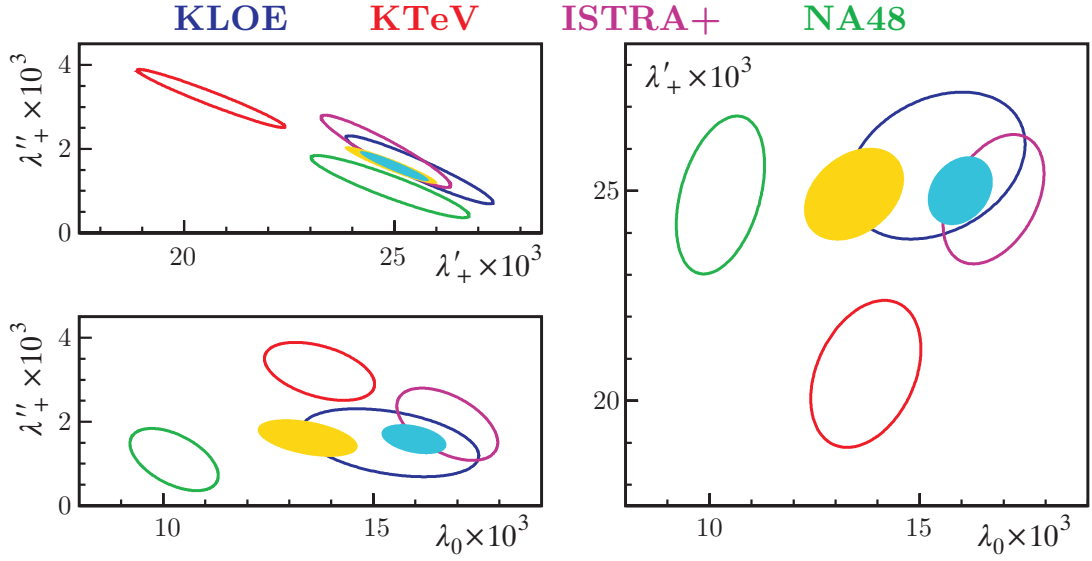}
\end{center}
\caption{\label{fig:FFm3} 1-$\sigma$ contours for $\lambda_+'$, $\lambda_+''$,
  $\lambda_0$ determinations from ISTRA+(pink ellipse), KLOE(blue ellipse),
  KTeV(red ellipse), NA48(green ellipse), and world 
 average with(filled yellow ellipse) and without(filled cyan ellipse) 
 the NA48 $K_{\mu3}$ result.}
\end{figure}

 Figure \ref{fig:FFm3} shows the 1-$\sigma$ contours from all
 the experimental results ($K_{e3}$ and $K_{\mu3}$). It is immediately
 clear from the figure that the new NA48 results are difficult to accommodate
\footnote{It lies out of correlation directions in the
 [$\lambda_+'$, $\lambda_+''$, $\lambda_0$] space}.
 Performing the combination with and without
 the NA48 results for the $K_{\mu3}$  form-factor slopes included
 we obtain fit probability values of \SN{1}{-6} and 22.3\%
 respectively(see Appendix~\ref{app:ff} for a detailed  comparison).
 The results of the combination are listed in \Tab{tab:l3ff}.

\TABLE{
\begin{tabular}{lcc}
\hline\hline
                                 & $K_L$ and $K^-$            & $K_L$ only     \\
\hline
Measurements                     & 16		              & 11	      \\
$\chi^2/{\rm ndf}$               & 54/13 $(7\times 10^{-7})$  & 33/8 $(8\times 10^{-5})$ \\
$\lambda_+'\times 10^3 $         & $24.9\pm1.1$ ($S=1.4$) & $24.0\pm1.5$ ($S=1.5$) \\
$\lambda_+'' \times 10^3 $       & $1.6\pm0.5$ ($S=1.3$)  & $2.0\pm0.6$ ($S=1.6$) \\
$\lambda_0\times 10^3 $          & $13.4\pm1.2$ ($S=1.9$)  & $11.7\pm1.2$ ($S=1.7$) \\
$\rho(\lambda_+',\lambda_+'')$   & $-0.94$                   & $-0.97$       \\
$\rho(\lambda_+',\lambda_0)$     & $+0.33$                   & $+0.72$       \\
$\rho(\lambda_+'',\lambda_0)$    & $-0.44$                   & $-0.70$       \\
$I(K^0_{e3})$                    & 0.15457(29)	      & 0.1544(4)  \\
$I(K^\pm_{e3})$                  & 0.15892(30)	      & 0.1587(4)  \\
$I(K^0_{\mu3})$                  & 0.10212(31)	      & 0.1016(4)  \\
$I(K^\pm_{\mu3})$                & 0.10507(32)	      & 0.1046(4)  \\
$\rho(I_{e3},I_{\mu3})$          & $+0.63$            & $+0.89$     \\
\hline\hline
\end{tabular}
\caption{Averages of quadratic fit results for $K_{e3}$ and $K_{\mu3}$ slopes.}
\label{tab:l3ff}
}

The value of $\chi^2/{\rm ndf}$ for all measurements is terrible;
we quote the results with scaled errors. This leads to
errors on the phase-space integrals that are $\sim$60\% larger after inclusion
of the new  $K_{\mu3}$  NA48 data.

We have checked to see if the NA48 $K_{\mu3}$
data might show good consistency with the results of some other
experiment in a less inclusive average.
Fitting to only the $K_{\mu3}$ results from KTeV, NA48, and ISTRA+ gives
$\chi^2/{\rm ndf} = 28/6$ (0.01\%).
Fitting to only the $K_{L\:\mu3}$ results from KTeV, NA48 gives
$\chi^2/{\rm ndf} = 12/3$ (0.89\%).
The consistency of the NA48 data with these other measurements appears
to be poor in any case.

The evaluations of the phase-space integrals for all four modes are listed in each case.
Correlations are fully accounted for, both in the fits and in the evaluation
of the integrals. The correlation matrices for the integrals are of the
form
\begin{displaymath}
\begin{array}{cccc}
+1 & +1 & \rho & \rho \\
+1 & +1 & \rho & \rho \\
\rho & \rho & +1 & +1 \\
\rho & \rho & +1 & +1 \\
\end{array}
\end{displaymath}
where the order of the rows and columns is $K^0_{e3}$, $K^\pm_{e3}$,
$K^0_{\mu3}$, $K^\pm_{\mu3}$,
and $\rho = \rho(I_{e3},I_{\mu3})$ as listed in the table.

Adding the $K_{\mu3}$ data to the fit does not cause drastic changes
to the values of the phase-space integrals for the $K_{e3}$ modes:
the values for $I(K^0_{e3})$ and $I(K^\pm_{e3})$ in \Tab{tab:l3ff}
are qualitatively in agreement with those in \Tab{tab:e3ff}.
As in the case of the fits to the $K_{e3}$ data only, the significance of the
quadratic term in the vector form factor is strong ($3.6\sigma$ from the
fit to all data).

\section{Physics Results}
\label{sec:results}

\subsection{Determination of $\vert V_{us}\vert\times f_{+}(0)$ and
 $\vert V_{us}\vert/\vert V_{ud}\vert\times f_K/f_\pi$}

 This section describes the results that are independent on the 
 theoretical parameters   $f_{+}(0)$ and $f_K/f_\pi$.

\subsubsection{Determination of $\vert V_{us}\vert\times f_{+}(0)$ }
 The value of $\vert V_{us}\vert\x f_{+}(0)$ has been determined from (\ref{eq:Mkl3}) using
 the world average values reported in section~\ref{sec:data}
 for lifetimes, branching ratios and phase space integrals, 
 and the radiative and $SU(2)$ breaking corrections
 discussed in section~\ref{sec:Hsu}.

\begin{table}[t]
\begin{center}
\begin{tabular}{l|c|c|c|c|c|c}
mode               & $\vert V_{us}\vert\x f_{+}(0)$  & \% err & BR   & $\tau$ & $\Delta$& Int \\
\hline
$K_L \to \pi e \nu$     & 0.2163(6)   & 0.28   & 0.09 & 0.19   & 0.15  & 0.09\\
$K_L \to \pi \mu \nu$   & 0.2168(7)   & 0.31   & 0.10 & 0.18   & 0.15  & 0.15\\
$K_S \to \pi e \nu$     & 0.2154(13)  & 0.67   & 0.65 & 0.03   & 0.15  & 0.09\\
$K^\pm \to \pi e \nu$   & 0.2173(8)   & 0.39   & 0.26 & 0.09   & 0.26  & 0.09\\
$K^\pm \to \pi \mu \nu$ & 0.2176(11)  & 0.51   & 0.40 & 0.09   & 0.26  & 0.15\\
 average & 0.2166(5)  &    &  &   &   & 
\end{tabular}
\end{center}
\caption{\label{tab:Vusf0}
 Summary of $\vert V_{us}\vert\x f_{+}(0)$ determination from all channels.}
\end{table}

\begin{figure}[t]
\centering
\includegraphics[width=0.4\textwidth]{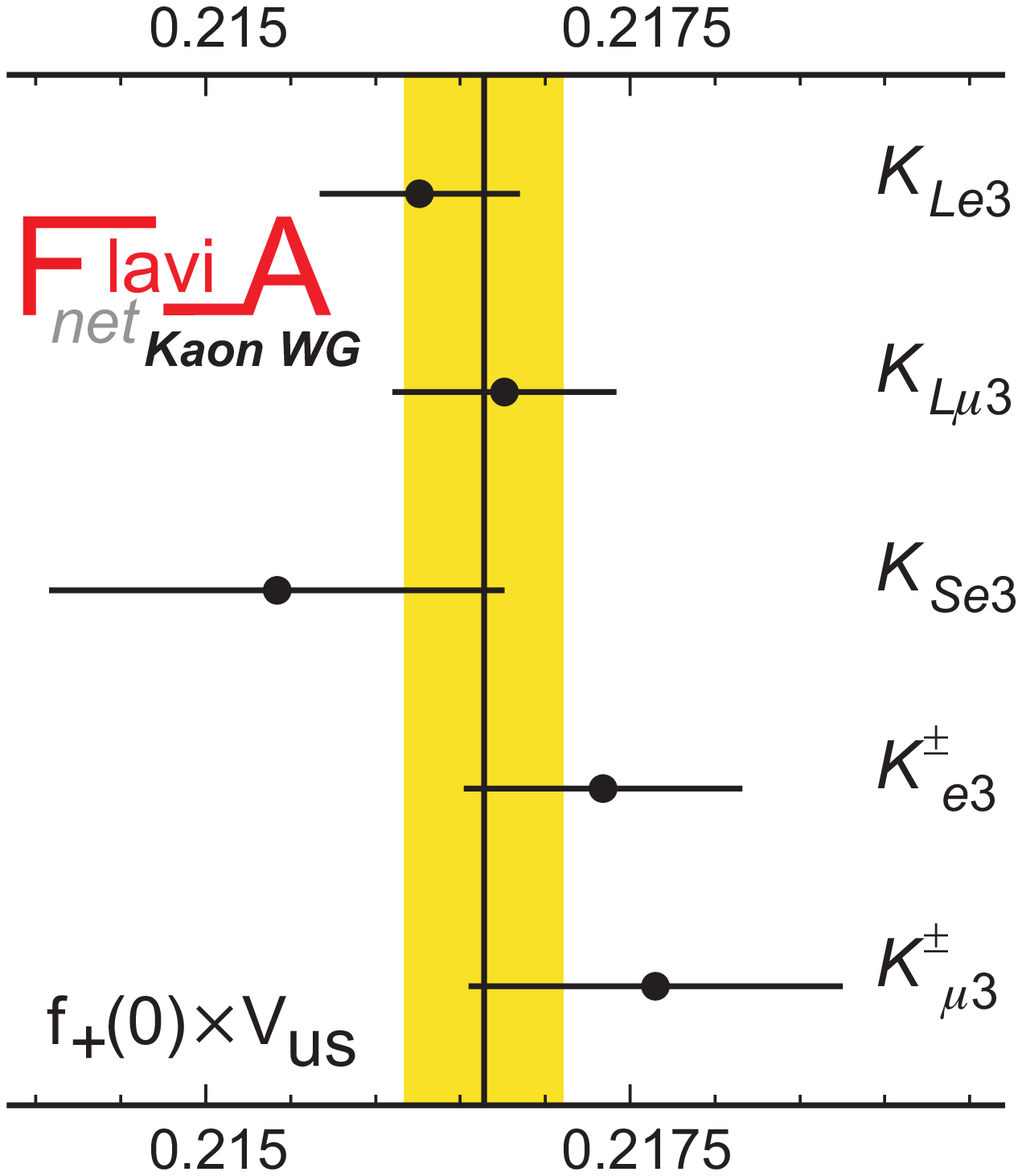}
\caption{ Display of $\vert V_{us}\vert\x f_{+}(0) $ for all channels. 
\label{fig:Vusf0} }
\end{figure}

The results  are given in Table~\ref{tab:Vusf0},
and are shown in \Fig{fig:Vusf0} for  $K_L \to \pi e \nu$, $K_L \to \pi \mu \nu$,
$K_S \to \pi e \nu$, $K^\pm \to \pi e \nu$, $K^\pm \to \pi \mu \nu$, 
and for the combination.
The average,   
\be
\vert V_{us}\vert\x f_+(0)=0.21664(48),
\label{eq:vusfres}
\ee
 has an uncertainty of about of $0.2\%$.  
The results from the five modes are in good agreement, the
fit probability is 58\%. 
In particular, comparing the values of $\vert V_{us}\vert\x f_{+}(0)$
obtained from $K^0_{\ell3}$ and $K^\pm_{\ell3}$ we obtain
a value of the SU(2) breaking correction 
$$
\delta^K_{SU(2)_{exp.}}= 2.9(4)\%
$$
in agreement
with the CHPT calculation reported in Table~\ref{tab:iso-brk}:
$\delta^K_{SU(2)}= 2.36(22)\%$.\footnote{~The value of $\delta^{K}_{SU(2)}$  
has a direct correspondence to the ratio of light quark masses. 
Recent analyzes~\cite{Donoghue-Perez}
on the so-called violations of Dashen's theorem in the Kaon 
electromagnetic mass splitting point to $\delta^{K}_{SU(2)}$ 
values of about $3\%$~\cite{kas}.}


\subsubsection{Determination of 
$\vert V_{us}\vert/\vert V_{ud}\vert\times f_K/f_\pi$ }
\label{sec:fkfpvusvud}

An independent  determination of \Vus\ is obtained from $K_{\ell2}$ decays. 
The most important mode is  $K^+\to\mu^+\nu$, which has been recently 
updated by KLOE reaching a relative uncertainty of about $0.3\%$.  
As shown in Eq.~(\ref{eq:Mkl2}), hadronic uncertainties are
minimized considering the ratio 
$\Gamma(K^+\to\mu^+\nu)/\Gamma(\pi^+\to\mu^+\nu)$.

Using the world average values
of BR($K^\pm\to\mu^\pm\nu$) and of $\tau^\pm$ given in Section~\ref{sec:data}
and the value of $\Gamma(\pi^\pm\to\mu^\pm\nu)=38.408(7)~\mu s^{-1}$
from \cite{bib:pdg} we obtain:
\beq
\vert V_{us}\vert/\vert V_{ud}\vert\x f_K/f_\pi = 0.2760 \pm  0.0006~.
\label{eq:vusvudres}
\eeq

\subsection{The parameters  $f_+(0)$ and $f_K/f_\pi$ }
\label{sec:ffzero}

The main obstacle in transforming these highly precise determinations of 
$\vert V_{us}\vert\x f_{+}(0)$ and $\vert V_{us}\vert/\vert V_{ud}\vert\times f_K/f_\pi$ into a determination of 
$\vert V_{us}\vert $ at the per-mil level are the theoretical 
uncertainties on the hadronic parameters $f_+(0)$ and $f_K/f_\pi$.

\subsubsection{Theoretical estimates of $f_+(0)$}
By construction, $\fp$  is defined in the absence of isospin-breaking effects
of both electromagnetic and quark-mass origin.
More explicitly, as discussed in Section~\ref{sect:KlSM},
$f_{+}(0)$ is defined 
by the $K^0 \to \pi^+$ matrix element of the vector current in the limit 
$m_u = m_d$ and $\alpha_{\rm em} \to 0$, keeping kaon and pion masses 
to their physical values.

This hadronic quantity cannot be computed in perturbative QCD, but
it is highly constrained by $SU(3)$ and chiral symmetry. 
In the chiral limit and, more generally, in the $SU(3)$ limit
($m_u=m_d=m_s$) the conservation of the vector current (CVC) 
implies $\fp$=1. Expanding around the chiral limit in powers 
of light quark masses we can write
\be
\label{chpt4}
f_+(0)= 1 + f_2 + f_4 + \ldots
\ee
where $f_2$ and $f_4$ are the NLO and 
NNLO corrections in ChPT.  The Ademollo--Gatto theorem implies that
$[f_+(0)-1]$ is at least of second order in the breaking of $SU(3)$ or in the 
expansion in powers of $m_s-\hat m$, where $\hat m=(m_u+m_d)/2$. This in turn implies 
that  $f_2$ is free from the uncertainties  of the $\cO(p^4)$ counterterms in ChPT, 
and it  can be computed with high accuracy: $f_2=-0.023$~\cite{LR}. 

\begin{figure}[t]
\centering
\includegraphics[width=0.7\textwidth]{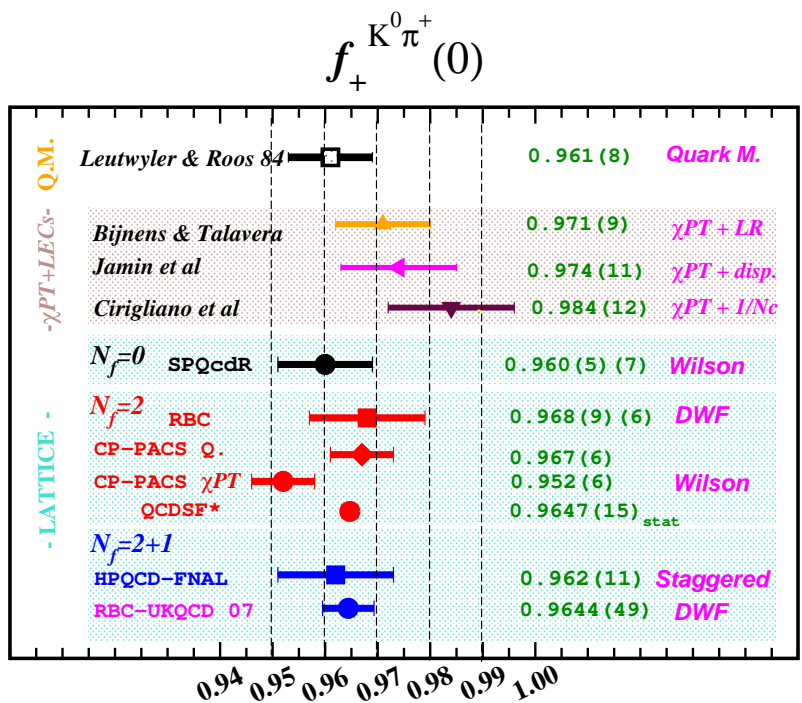}
\caption{\label{f0}Present determinations 
of $f_+(0)\equiv f_+^{K^0\pi^-}(0)$ from lattice
QCD and analytical or semi-analytical approaches~\cite{LR,rbcf0,milcf0,f0}.  }
\end{figure}

The difficulties in 
estimating $f_+(0)$ begin with $f_4$ or at $\cO(p^6)$ in the chiral expansion.
At this order we can write 
\be 
\renewcommand{\arraystretch}{0.5} 
f_4 = \Delta(\mu) + f_4\vert^{\rm loc}(\mu)\,, 
\label{eq:f4ch}
\ee 
where $\Delta(\mu)$ is the loop contribution, which has been computed 
in Ref.~\cite{Bijnens:2003}, and $f_4\vert^{\rm loc}(\mu)$ is the 
$\cO(p^6)$ local contribution, whose knowledge cannot be simply 
deduced from other processes. Several analytical approaches to determine $f_4$
have been attempted over the years~\cite{f0}, essentially confirming the original
estimate by Leutwyler and Roos~\cite{LR} (see \Fig{f0}).
The benefit of these new results, obtained using 
more sophisticated techniques, lies in the fact that a 
better control over the systematic uncertainties of the calculation
has been obtained. However, the size of the error is still around or 
above $1\%$, which is not comparable  to the $0.2\%$
accuracy which has been reached for $\vert V_{us}\vert\x f_+(0)$.

\begin{figure}[t]
\centering
\includegraphics[width=0.8\textwidth]{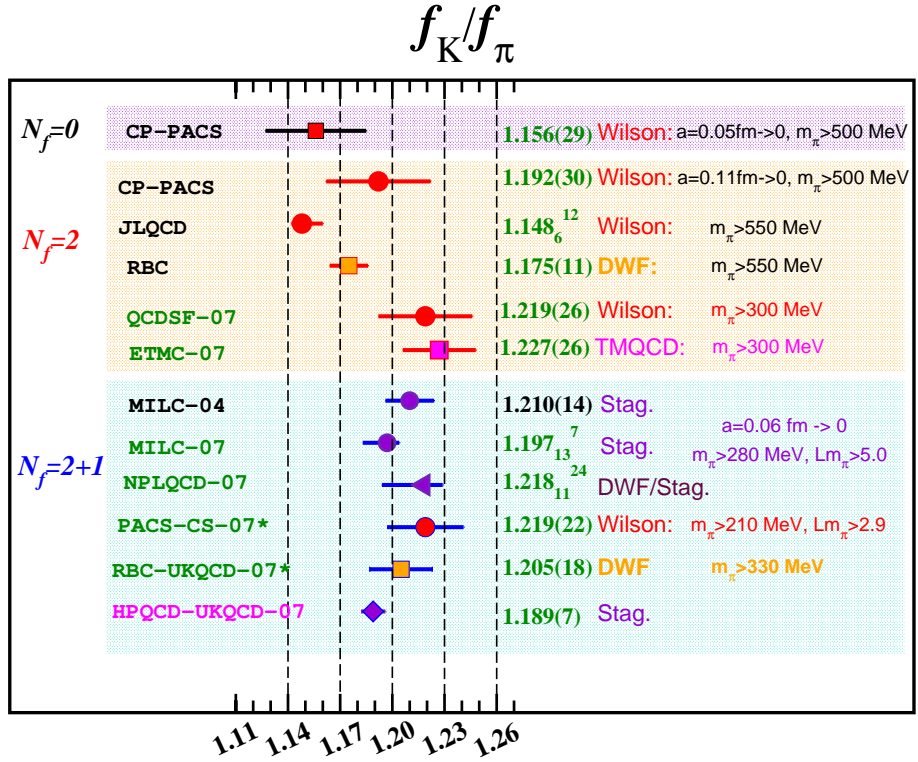}
\caption{\label{fk}Summary of $f_K/f_\pi$
estimates~\cite{milcfk,davies,rbcfk,fk}. All values are from Lattice QCD.               
In recent studies, sea quarks are getting light and data are  
matched to ChPT fits to determine the
Low-energy-Constants (LEC). }
\end{figure}

Recent progress in lattice QCD gives us more optimism in the reduction of 
the error on $\fp$ below the $1\%$ level~\cite{kan,jut,f0latt}.  
Most of the currently available 
lattice QCD  results have been obtained with relatively heavy pions and  
the chiral extrapolation represents the dominant source of uncertainty.
As shown in  Figure~\ref{f0}, there is a general trend of 
 lattice QCD results to be slightly lower than  analytical approaches. 
An important step in the reduction of the error associated to the 
chiral extrapolation  has been recently made by 
the UKQCD-RBC collaboration~\cite{rbcf0}. 
Their preliminary  result $\fp=0.964(5)$
is obtained from the unquenched study  with 
$N_F=2+1$ flavors, with an action that has good chiral properties 
on the lattice even at finite lattice 
spacing (domain-wall quarks). They also reached pions masses ($\gtrsim 330\,\mev$) 
much lighter than that used in  previous studies of $f_+(0)$. The 
overall error is estimated to be  $~0.5\%$, which is very encouraging.
Moreover, they observe for $f_+(0)$
a mass dependence similar to the one of $f_2$. That is something new 
with respect to previous lattice studies (this is likely
due to the fact that they work with lighter pions). To  assess
the chiral uncertainty of $f_4$, polynomial fits 
(linear and quadratic) well reproduce the data. However,  
it would be interesting~\cite{jut} to have the expression
of $\Delta(\mu)$ in Eq.~(\ref{eq:f4ch}) in terms of the quark masses so to
directly estimate $f_4\vert^{\rm loc}(\mu)$. 
Moreover, it should also be stressed that the present study is performed at 
a single value of the lattice spacing ($a=0.12$ fm) and 
in a relatively small 
extension of the fifth dimension of the lattice.\footnote{~Even though $m_\pi L\gtrsim 4.5$, 
 simulations with a larger fifth dimension, $L_s$ would help too because  
the mass of their lightest quark ($=0.005$ in lattice units)  is very close to the
 residual mass parameter ($=0.003$, also in lattice units). 
 This may entail some uncontrolled
 systematics, in particular for  $f_K/f_\pi$}
 
In the following phenomenological analysis we will use this result as
the present best estimate of  $f_+(0)$, although some reservation remains.

\subsubsection{Theoretical estimates of $f_K/f_\pi$}

In contrast to the semileptonic vector form factor, the pseudoscalar
decay constants are not protected by the Ademollo--Gatto theorem and 
receive corrections linear in the quark masses. Expanding 
$f_K/f_\pi$ in power of quark masses, in analogy to $f_+(0)$, 
\be
\label{chpt3}
f_K/f_\pi= 1 + r_2 + \ldots
\ee
one finds that the $\cO(p^4)$ contribution $r_2$ is 
already affected by local contributions and cannot be unambiguously 
predicted in ChPT. As a result, in the determination of $f_K/f_\pi$
lattice QCD~\cite{milcfk}-\cite{fk} has 
essentially no competition from purely analytical approaches. The status of the 
lattice results for $f_K/f_\pi$ is  summarized 
in Fig.~\ref{fk}. As can be seen, the  present overall accuracy is about $1\%$. 
The  novelty are the new  lattice results with 
 $N_F=2+1$ dynamical quarks  and  pions as light as  $280$~MeV~\cite{milcfk,davies},
 obtained by using the so-called staggered quarks.\footnote{~Staggered 
fermions come in four tastes on the lattice. In the continuum limit  
 the extra degrees of freedom decouple from
 physical predictions.  But, at finite
 lattice spacing, where the data are produced, the taste symmetry is violated 
 and the extra degrees of freedom are removed by hand, 
namely by taking the fourth root of the staggered
 quark determinant. Theoretically, this procedure has been 
 only confirmed in perturbation theory and is currently a subject of controversies 
 within the lattice QCD community~\cite{kronfeld-creutz}. 
Since the staggered dynamical quarks are computationally cheap, they have been largely used. Thanks to 
  recent progress in algorithm building~\cite{algo},
  safer but still computationally competitive alternatives 
  are becoming available.
 } 
The analyzes of \cite{milcfk,davies},
cover a broad range of lattice spacings (i.e.~$a$=0.06 and 0.15 fm) and  
is performed on sufficiently large physical volumes ($m_\pi L\gtrsim 5.0$). 
It should be stressed, however, that the sensitivity of 
$f_K/f_\pi$ to lighter pions is larger 
than in the computation of $\fp$ and that  chiral extrapolations are far more 
demanding in this case.\footnote{~In some details,  
effects of chiral logs are not clearly disentangled and
analytic terms (NNLO or NNNLO) are still needed in order to extrapolate from the
simulated sea
quark masses (such as $m_\pi\gtrsim 280$ MeV) to the physical point. 
For example, the two studies of ref.~\cite{milcfk} and of ref.~\cite{davies} with staggered quarks
share the same configurations, but they differ in how to extrapolate to the physical masses.
Then, the central values of $f_K/f_\pi$ between the two analyzes 
 (namely, $f_K/f_\pi=1.197^7_{13}$ and $f_K/f_\pi=1.189(7)$ from ref.~\cite{milcfk} and ref.~\cite{davies}
respectively) differ for $1\sigma$. However, taking into account the complete uncertainty
 of $f_K/f_\pi$ in~\cite{milcfk}, 
we have $f_K/f_\pi=1.194(10)$ of~\cite{milcfk} for 
a symmetric error and the values of ref.~\cite{milcfk} and ref.~\cite{davies} look now 
in good agreement. The highly improved staggered fermions
  (HISQ) used in~\cite{davies} for the valence quarks are
  designed to reduce the taste violation effects, which also should
reduce the overall systematic uncertainty. }
Notice also that at Lattice 2007
   preliminary studies with $N_F=2+1$  clover quarks and pion masses 
   $\gtrsim 200$ MeV have been presented 
from either   PACS-CS Collaboration~\cite{jam} and 
ref.~\cite{lel}.  
   With respect to the results obtained with staggered
quarks, the PACS-CS  value of $f_K/f_\pi$ in fig.~\ref{fk} is 
restricted to a single lattice spacing ($a=0.09$ fm) and 
relatively small physical
volume ($m_\pi L\gtrsim 2.9$). For ref.~\cite{lel},
  the final analysis is to be completed.
 In the following 
   analysis we will use as reference value the MILC-HPQCD
   result  $f_K/f_\pi=1.189(7)$~\cite{davies}, 
    although some reservation about staggered fermions remains.

\subsubsection{A test of lattice calculation: the Callan-Treiman relation}
\label{sec:CTtest}
 As described in Sect.~\ref{sec:ffpara} the Callan-Treiman relation fixes the
 value of scalar form factor at $t=m_K^2-m_\pi^2$ 
(the so-called Callan-Treiman point) to the
 ratio $(f_K/f_\pi)/f_+(0)$. The dispersive parametrization for the scalar
 form factor proposed in~\cite{stern} and discussed in Sect.~\ref{sec:ffpara}
 allows to transform the available measurements of the scalar form factor 
 into a precise information on $(f_K/f_\pi)/f_+(0)$, completely independent of
 the lattice estimates.

 Very recently KLOE~\cite{KLOE+07:m3FF} 
 and NA48~\cite{NA48+06:m3FF} have presented results on the
 scalar FF slope  using the dispersive parametrization.
  In these analyzes a dispersive 
 parametrization is used for both the scalar and the vector form factors. 
 A similar analysis has started for the KTeV data.
 We report these  preliminary results for the first time.
 The  ISTRA+ measurement of the scalar
 form factor slope performed using the first order Taylor expansion
 parametrization can be translated in the dispersive parametrization
 as described in Appendix~\ref{app:Fslope}.
 The results are given in Table~\ref{tab:ffdis} for all the four experiments in the case
 of the pole parametrization for the vector form factor.
 The original KLOE and NA48 results are also shown for comparison as well as the 
 preliminary result of KTeV obtained from the $K_{\mu3}$ data analysis. 
 Moreover, a combined $K_{e3}$ and $K_{\mu3}$ data analysis is also in progress and   
 the preliminary result is: $\log(C)=0.191 \pm 0.012$.  
 The preliminary KTeV results are obtained using the original 
 MC and data from Ref. \cite{KTeV+04:FF}. .

\begin{table}[ht]
\centering
\begin{tabular}{l|c|c}
Experiment & $\log(C)$ direct & $\log(C)^\dag$\\
\hline
KTeV   & 0.195(14)$^*$  & 0.203(15)   \\
KLOE   & 0.207(24)      & 0.207(23)   \\
NA48   & 0.144(14)      & 0.144(13)   \\
ISTRA+ &                & 0.226(13)   \\
\hline
\end{tabular}\\[1mm]
$^\dag$ {\small Estimated from $\lambda_0$ published.}$^*${\small Preliminary results.}\kern2cm\\
\caption{\label{tab:ffdis}
Experimental results for log(C).}
\end{table}


\begin{figure}[t]
\centering
\includegraphics[width=0.5\textwidth]{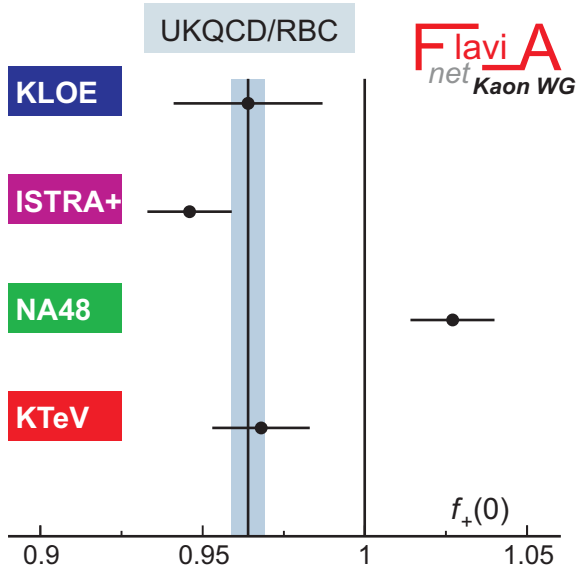}
\caption{ Values for $f_+(0)$ determined from the scalar form factor slope using the Callan-Treiman 
relation and  $f_K/f_\pi=1.189(7)$. The UKQCD/RBC result $f_+(0)=0.964(5)$ is also
shown.\label{fig:CTtest} }
\end{figure}

Figure \ref{fig:CTtest} shows the values for $f_+(0)$ 
determined from the scalar form factor slope
measurements obtained using the Callan-Treiman relation and
$f_K/f_\pi=1.189(7)$. The value of $f_+(0)=0.964(5)$ from UKQCD/RBC is also shown.
As already noticed in Section~\ref{sec:data}, the NA48 result is 
difficult to accommodate. Here one can see that 
this results is also not consistent with the theoretical 
estimates of $f_+(0)$. In particular, it violates the 
Fubini-Furlan bound $f_+(0)<1$~\cite{furlan}. For this
reason, the NA48 result will be excluded when using the
Callan-Treiman constraint.

The average of the experimental results on the FFs with the
pole parametrization for the vector case and the dispersive parametrization
for the scalar FF give:
\be
\eqalign{
  \lambda_+^c&=0.0256\pm 0.0002~,\cr
  \lambda_0^c&=0.0149\pm 0.0007~,\cr}
\ee
with correlation coefficient \minus0.32.  The above results are then 
 combined with the lattice determinations of  $f_K/f_\pi=1.189(7)$ and 
 $f_+(0)=0.964(5)$ using the constraint given by the Callan-Treiman
 relation. The results of the combination are given in Table~\ref{tab:ffsct},
where $\log C= \lambda_0^c~t_{\rm CT}/m^2_\pi + 0.0398 \pm 0.0041$.
 
\begin{table}[ht]
\begin{center}
\begin{tabular}{c|c|c|c}
$\lambda_+^c$ & $\lambda_0^c$ & $f_+(0)$ & $f_K/f_\pi$ \\
\hline
  0.02563(19) & 0.0146(5) & 0.96(4) & 1.192(6)\\
\hline
\multicolumn{4}{c}{correlation matrix}\\
\hline
          1.    & -0.23  &  0.12 & -0.14\\
                &  1.    & -0.51 &  0.61\\
                &        &  1.   &  0.30\\
                &        &       &  1.\\
\end{tabular}
\end{center}
\caption{\label{tab:ffsct}Results from the form factor fit.}
\end{table}
 
 The fit probability is 39\%, confirming the agreement between
 experimental measurements and lattice determination.
 The accuracy of $f_K/f_\pi$ is also slightly improved, and 
 this effect can be better seen in the ratio  $f_+(0)/(f_K/f_\pi)$,
 directly related to the Callan-Treiman constraint.

As previously discussed, new physics contributions to the 
scalar form factor (reabsorbed into the value of $\log C$) 
are generated only by scalar operators. Hence in the case of 
right-handed currents $\log C$ coincides with the SM value.
This imply we can use the Callan-Treiman improved  $f_+(0)/(f_K/f_\pi)$
in constraining right-handed currents. On the other hand,
this is not possible in the MSSM scenario, where
scalar operators are present. Here the measured value of $\log C$,
following from (\ref{eq:f0mssm}), is 
 \be 
 \log C^{\rm MSSM} \equiv \log\left\vert f^H_0(t_{CT})/f_0(0)\right\vert= 
 \log C^{\rm SM} - r^K_H
 \ee
with the $r^K_H$ given in~(\ref{eq:rk}). By construction, the quantity
$\log C^{\rm SM}$ depends only on QCD dynamics and must satisfies 
the Callan-Treiman relation (\ref{eq:CTrel}). 
The theoretical  calculation of
 $f_0(t_{CT})$ can thus be used 
 to constrain scalar densities. 
 At present, the theoretical knowledge of $\log C^{\rm SM}$ is 
obtained from Eq.~(\ref{eq:CTrel}) and is limited  
by our knowledge of $\Delta_{CT}$, reported in (\ref{eq:DeltaCT}), 
and by the lattice QCD results on $(f_K/f_\pi)/f_+(0)$.
Using this information we obtain the constraint
\be
r^K_H = -0.007\pm0.012~.
\ee
To improve this result it would be particularly useful a 
direct computation of $(f_K/f_\pi)/f_+(0)$ on the lattice (i.e.~from the 
the same set of simulations). Given the advanced status of 
staggered results on $f_K/f_\pi$, it would be interesting 
to see the effect of a corresponding analysis $f_+(0)$
(which at present is still very preliminary~\cite{milcf0}). 

\begin{figure}[t]
\centering
\includegraphics[width=0.6\textwidth]{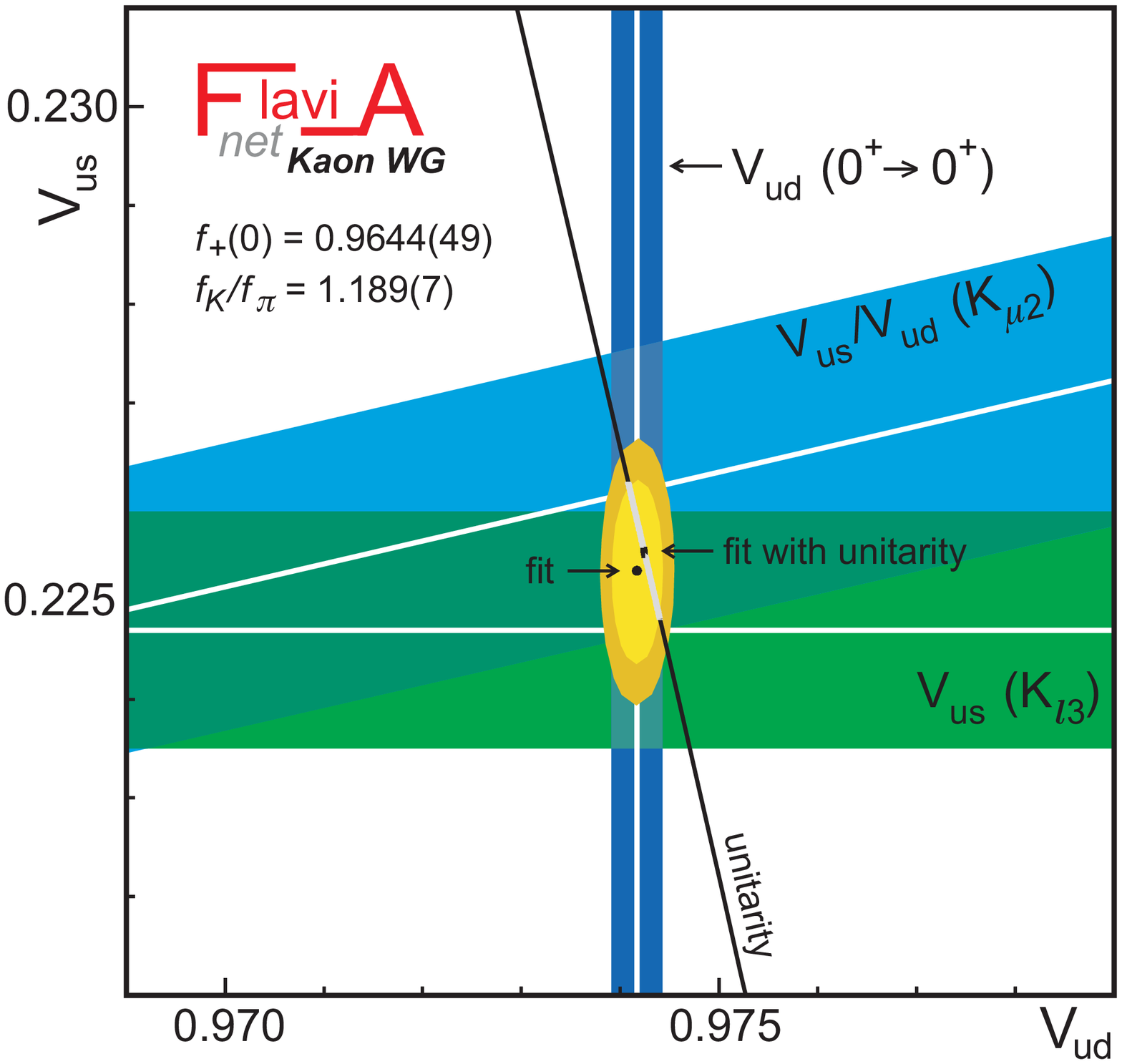}
\caption{\label{fig:vusuni} Results of fits to \Vud, \Vus, and $\Vus/\Vud$.}
\end{figure}

\subsection{Test of Cabibbo Universality or CKM unitarity}
To determine $|V_{us}|$ and   $|V_{ud}|$
 we use the value $\vert V_{us}\vert\x f_{+}(0)=0.2166(5)$ reported 
in  Table~\ref{tab:Vusf0}, the result $\vert V_{us}\vert/\vert V_{ud}\vert f_K/f_\pi = 0.2760(6)$ 
discussed in Sect.~\ref{sec:fkfpvusvud}, $f_+(0) = 0.964(5)$, and $f_K/f_\pi = 1.189(7)$. 
From the above we find:
\bea
 \vert V_{us}\vert&=& 0.2246\pm  0.0012 \qquad [K_{\ell 3}~{\rm only}]~, \\
 \vert V_{us}\vert/\vert V_{ud}\vert&=& 0.2321\pm  0.0015 \qquad [K_{\ell 2}~{\rm only}]~.
\eea
 These determinations can be used in a fit together with the 
 the recent evaluation of \Vud\ from
 $0^+\to0^+$ nuclear beta decays: $|V_{ud}|$=0.97418\plm0.00026~\cite{t&h}.
The global fit gives 
\be
\Vud = 0.97417(26) \qquad \Vus = 0.2253(9) \qquad [K_{\ell 3,\ell 2}~+~ 0^+\to0^+]~,
\ee
with $\chi^2/{\rm ndf} = 0.65/1$ (42\%). This result does not make use 
of CKM unitarity. If the  unitarity constraint is included, 
the fit gives 
\beq
\Vus=\sin\,\theta_C=\lambda=0.2255(7) \qquad [{\rm with~unitarity}]
\eeq
and $\chi^2/{\rm ndf}=0.80/2$ (67\%).
Both results are illustrated in \Fig{fig:vusuni}.

\begin{figure}[t]
\centering
\includegraphics[width=0.7\textwidth]{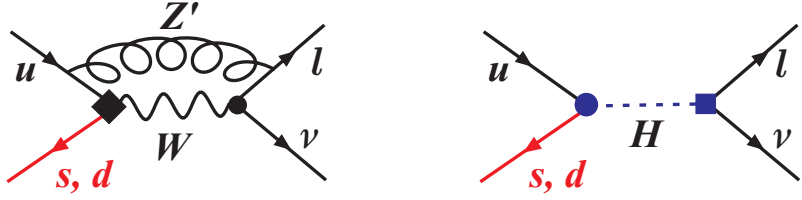}
\caption{\label{fig:zpgraph} $Z^\prime$ and Higgs exchange.}
\end{figure}

 As described in the introduction, the test of CKM unitarity
 can be also interpreted as a test of universality of
 the lepton and quark gauge couplings.
 Using the results of the fit (without imposing unitarity) we obtain:
\begin{equation}
G_{\rm CKM} \equiv G_\mu \left[ |V_{ud}|^2+|V_{us}|^2+|V_{ub}|^2 \right]^{1/2}
= (1.1662 \pm  0.0004)\times 10^{-5}\  {\rm GeV}^{-2}~,
\end{equation}
in perfect agreement with the value obtained from the measurement
of the muon lifetime:
\begin{equation}
G_{\mu} = (1.166371 \pm  0.000007)\times 10^{-5}\  {\rm GeV}^{-2}\,.
\end{equation}
 The current accuracy of the lepton-quark universality
 sets important constraints on model building beyond the SM.
 For example, the presence of  a $Z^\prime$ (see Fig.~\ref{fig:zpgraph}, left) 
 would affect the relation between $G_{\rm CKM}$ and $G_{\mu}$
 in the following way,
\begin{equation}
G_\mu=G_{CKM}
\left[ 1-0.007
Q_{eL}(Q_{{\mu}L}-Q_{dL}) \frac{ 2 \ln(m_{Z^\prime}/m_W)}{m_{Z^\prime}^2/m_W^2-1}\right]~,
\label{eq:zprimo}
\end{equation}
where $Q_{fL}$ are the generic charges of the $Z^\prime$ to left-handed 
leptons (in units of the SM $SU(2)_L$ charge). 
In case of a $Z^\prime$ from $SO(10)$ grand unification theories
($Q_{eL}=Q_{{\mu}L}=-3Q_{dL}=1$) we obtain  $m_{Z^\prime}>700$~GeV at 95\% CL, 
to be compared with the  $m_{Z^\prime}>720$~GeV
bound  set through the direct collider searches~\cite{bib:pdg}.
In a similar way, the 
unitarity constraint also provides useful bounds in various 
supersymmetry-breaking scenarios~\cite{barb}.

\subsubsection{Bounds on helicity-suppressed amplitudes}

A particularly interesting test is the comparison of the $\vert V_{us}\vert$ 
value extracted from the helicity-suppressed $K_{\ell 2}$ decays
with respect to the value extracted from the  helicity-allowed $K_{\ell 3}$  modes.
To reduce theoretical uncertainties from $f_K$ and electromagnetic 
corrections in $K_{\ell 2}$,
we exploit the ratio $Br(K_{\ell2})/Br(\pi_{\ell2})$ and 
we study the quantity
\be
R_{l23}=\left\vert \frac{V_{us}(K_{\ell 2})}{V_{us}(K_{\ell 3})}\x 
\frac{V_{ud}(0^+\to 0^+)}{V_{ud}(\pi_{\ell 2})}\right\vert\,.
\ee
Within the SM, 
 $R_{l23}=1$, while deviation from 1 can be induced by 
 non-vanishing scalar- or  right-handed currents.
 Notice that in $R_{l23}$ the  hadronic uncertainties
enter through  $(f_K/f_\pi)/f_+(0)$.

Following the notation of Section~\ref{sect:NP}, 
effects of scalar currents due to a charged Higgs (Fig.~\ref{fig:zpgraph} right)
 give 
 \be
 R_{l23} = \left\vert 1 - \frac{m^2_{K^+}}{M^2_{H^+}}\left(1-\frac{m_d}{m_s} \right)
\frac{\tan^2\beta}{1+\epsilon_0\tan\beta}\right\vert~,
\ee
whereas  for right-handed currents we have
 \be
 R_{l23} = 1 -2\,\left(\epsilon_s-\epsilon_{ns}\right)~.
\ee

In the case of scalar densities (MSSM),
the unitarity relation between 
$\vert V_{ud}\vert$ extracted from 
  $0^+\to0^+$ nuclear beta decays and $\vert V_{us}\vert$ extracted from 
$K_{\ell3}$ remains valid as soon as form factors are experimentally determined.
This constrain  together with the experimental
information of $\log C^{MSSM}$ 
can be used in the global fit to improve the accuracy of the determination 
of $R_{l23}$, which in this scenario turns to be 
\be
\left. R_{l23} \right|^{\rm exp}_{\rm scalar}  =  1.004 \pm  0.007~.
\ee
Here $(f_K/f_\pi)/f_+(0)$ has been fixed from lattice. This ratio
is the key quantity to be improved in order to reduce 
present uncertainty on $R_{l23}$. 

\begin{figure}[t]
\centering
\includegraphics[width=0.6\textwidth]{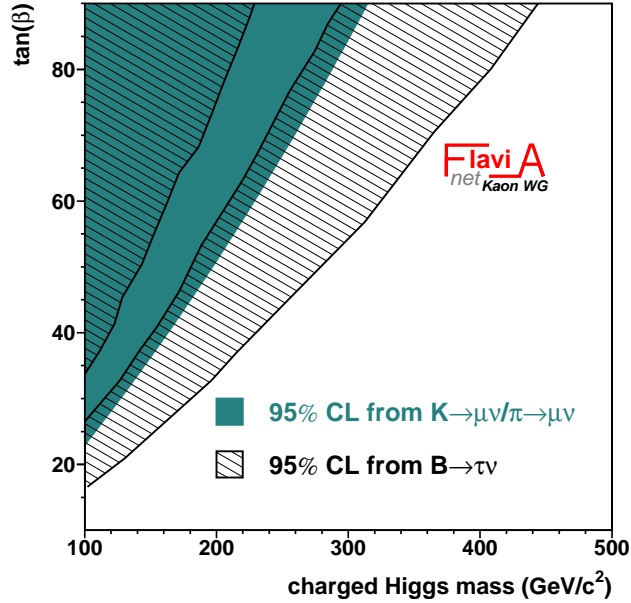}
\caption{\label{fig:higgskmunu}
Excluded region in the charged Higgs mass-$\tan\beta$ plane.
The region excluded by  $B\to \tau \nu $ is also indicated.}
\end{figure}

The  measurement of $R_{l23}$ above can be used to set bounds
on the charged Higgs mass and $\tan\beta$.
Figure \ref{fig:higgskmunu} shows the excluded region at 95\%
 CL in the $M_H$--$\tan\beta$ plane (setting $\epsilon_0=0.01$).
 The measurement of  BR($B \to \tau \nu$)~\cite{bib:btaunu}
 can be also used to set a similar  bound
in the  $M_H$--$\tan\beta$  plane. While $B\to\tau \nu$
 can exclude quite an extensive region of this plane,
 there is an uncovered region in the exclusion corresponding 
 to a destructive interference between the charged-Higgs 
 and the SM amplitude.
 This region is fully covered by the $K\to \mu \nu$ result.

In the case of right-handed currents~\cite{stern},
$R_{l23}$ can be obtained from a global fit to the values
of eqs.~(\ref{eq:vusfres}) and (\ref{eq:vusvudres}). 
Here $\log C^{\rm exp}$ is free of new physics effects and 
can be also used to constrain $(f_K/f_\pi)/f_+(0)$ 
together with lattice results (namely the values in tab.~\ref{tab:ffsct}).
The result is
\be
\left. R_{l23} \right|^{\rm exp}_{\rm RH curr.} =  1.004 \pm  0.006~. 
\ee

\subsection{Tests of Lepton Flavor Universality}
\subsubsection{Lepton universality in $K_{\ell 3}$ decays}
The test of  Lepton Flavor Universality (LFU) between  
$K_{e3}$ and $K_{\mu3}$ modes constraints a possible
anomalous lepton-flavor dependence in the leading 
weak vector current. It can therefore be compared
to similar tests in $\tau$ decays, but is different from the 
LFU tests in the helicity-suppressed modes $\pi_{l2}$ and $K_{l2}$.

The results on the parameter
 $r_{\mu e} = 
R_{K_{\mu3}/K_{e3}}^{\rm{Exp}}/R_{K_{\mu3}/K_{e3}}^{\rm{SM}}$ is
\be
r_{\mu e} = 1.004 \pm 0.004~,
\ee
in excellent agreement with lepton universality. 
Furthermore, with a precision of $0.5\%$ the test in $K_{l3}$ decays 
has now reached the sensitivity of $\tau$ decays.

\subsubsection{Lepton universality tests in $K_{\ell 2}$ decays}
The ratio $R_K = \Gamma({K_{\mu2}})/\Gamma({K_{e2}})$ can be precisely calculated 
within the
 Standard Model.
 Neglecting radiative corrections, it is given by 
\be
R_K^{(0)} = \frac{m_e^2}{m_\mu^2} \: \frac{(m_K^2 - m_e^2)^2}{(m_K^2 - m_\mu^2)^2} = 2.569 \times 10^{-5},
\ee
and reflects the strong helicity suppression of the electron channel.
Radiative corrections have been computed with effective
 theories~\cite{ciriglianokl2},
yielding the final SM prediction
\bea
R^{\rm SM}_K &=& R_K^{(0)} ( 1 + \delta R_K^{\text{rad.corr.}})  \nn \\
&=& 2.569 \times 10^{-5} \times ( 0.9622 \pm 0.0004 ) =(2.477 \pm 0.001) \times 10^{-5}~. 
\eea

Because of the helicity suppression within then SM, the 
 $K_{e2}$  amplitude is a prominent candidate
 for possible sizable contributions from physics beyond the SM. Moreover,
 when normalizing to the $K_{\mu2}$ rate, we obtain an extremely precise 
 prediction of the $K_{e2}$ width within the SM. In order to be visible 
 in the $K_{e2}/K_{\mu2}$ ratio, the new physics must  violate lepton 
 flavor universality.

 Recently it has been pointed out that in a supersymmetric framework 
sizable violations of  lepton universality can be expected
 in $K_{l2}$ decays~\cite{paride}. At the tree level, 
 lepton flavor violating terms are forbidden in the MSSM. 
 However, these appear at the one-loop level, where an effective 
 $H^+ l \nu_\tau$ Yukawa interaction is generated. Following the
 notation of Ref.~\cite{paride} (see also Section~\ref{sect:NP}),
 the non-SM contribution to $R_K$ can be written as 
\be
R_K^{\text{LFV}} \approx R_K^{\text{SM}} \left[ 1 + \left(
\frac{m_K^4}{M_{H^\pm}^4} \right) \left( \frac{m_\tau^2}{m_e^2} \right) |\Delta_{13}|^2 \tan^6 \beta \right]~.
\label{eqn:susy}
\ee
 The lepton flavor violating coupling $\Delta_{13}$, being generated at the 
loop level, could reach values of $\cO(10^{-3})$.
 For moderately large $\tan \beta$ values, 
 this contribution may therefore
 enhance $R_K$ by up to a few percent.
 Since the additional term in Eq.~\ref{eqn:susy} goes with the forth power of the
 meson mass, no similar effect is expected in $\pi_{l2}$ decays.

\begin{figure}[t]
\centering
\includegraphics[width=0.6\linewidth]{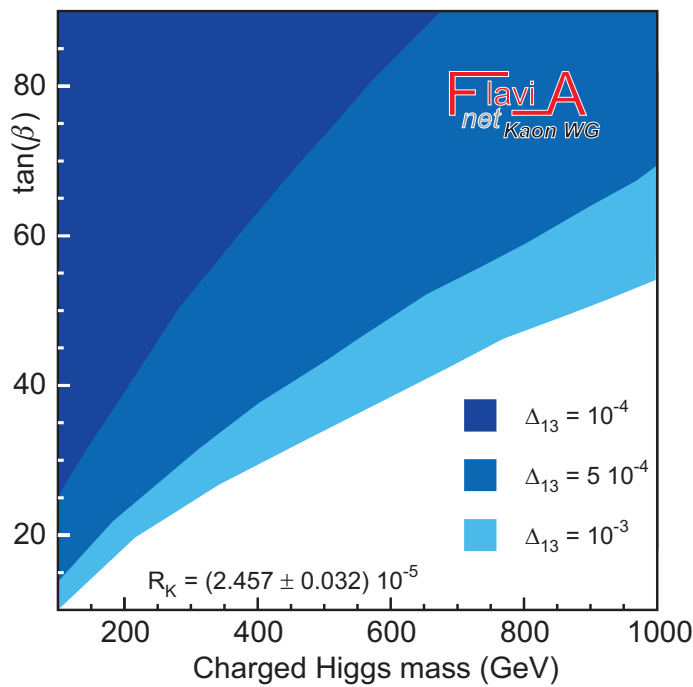}
\caption{Exclusion limits at $95\%$ CL on $\tan \beta$ and the charged
Higgs mass $M_{H^\pm}$ 
from $\vert V_{us}\vert_{K\ell2}/\vert V_{us}\vert_{K\ell3}$ for different
values of $\Delta_{13}$. }
\label{fig:susylimit}
\end{figure}

 The world average result for $R_K$ presented in Section \ref{sec:data} 
 gives strong constraints 
 for $\tan \beta$ and $M_{H^\pm}$, as shown in Fig.~\ref{fig:susylimit}.
 For values of $\Delta_{13} \approx 5 \times 10^{-4}$
 and  $\tan \beta > 50$ the charged Higgs masses is pushed 
 above 1000~GeV/$c^2$ at 95\% CL.


\addcontentsline{toc}{section}{Acknowledgments}

\section*{Acknowledgments}
We thank all the members of the FlaviaNet Kaon Working
Group  [{\footnotesize{\tt www.lnf.infn.it/wg/vus}}],
and in particular J.~Gasser and J.~Stern, 
for comments, discussions, and suggestions.
This work is  supported in part by the
EU contract No.~MTRN-CT-2006-035482 (FlaviaNet).


\appendix

\section{BRS fit procedure}
\label{app:fitprocedure}
The fits to $K_L$ and $K^\pm$ data are performed with \textsc{fortran} 
programs. \textsc{migrad} is used for the minimization; errors are 
obtained with \textsc{minos}.

Suppose we have $N$ measurements of $M$ quantities, e.g., BRs, ratios of BRs, 
lifetimes, or partial widths, where $N \geq M$ as some quantities are 
measured by more than one experiment.
Denote the $N$ measurements $x_i$, and the expected value for each 
as calculated from 
the free parameters of the fit $\bar{x}_i$.
We also refer to the expected values for quantities measured by more than 
one experiment by the index $m$, i.e., $\bar{x}_m$ with $m=1,M$.

The errors on the input parameters are denoted $\sigma_i$.
All errors on the input parameters are assumed to be Gaussian.
For uncorrelated measurements with statistical and systematic errors 
quoted separately, we add the errors in quadrature.
In many cases, the results for different quantities measured by 
the same experiment have correlated errors. The errors are then 
described by the covariance matrix $V_{ij}$, with $V_{ii} = \sigma^2_i$
and $V_{ij} = \rho_{ij}\sigma_i\sigma_j$. 
The expression to be minimized is then
\begin{equation}
\chi^2 = \sum_{i=1}^{N}\sum_{j=1}^{N}
(x_i - \bar{x}_i)(x_j - \bar{x}_j)(\vec{V})^{-1}_{ij}.
\end{equation}
In practice, \vec{V} is block diagonal and only the relevant sub-matrices 
are inverted.

The penalty method is used to implement the constraint on the sum of the 
BRs. In this method, a term $G\,(1-\sum{\rm BR})^2$ is added to the $\chi^2$ 
to be minimized. As $G$ is increased, the constraint is enforced with
greater and greater precision and the result of the fit saturates 
(until at some very large value of $G$, problems related to the precision
of the calculation set in). $G$ is determined by trial and error; 
its value is \SN{2}{7} for the $K_L$ fit and \SN{1}{8} for the $K^\pm$ fit.
The $K_L$ fit is somewhat more sensitive to the value of $G$, 
because the $K_L$ BRs entering the fit span three orders of magnitude.
As a result, precision problems have a greater effect on the constraint 
balance.

Once the fit has been performed, scale factors are calculated and used 
as described in the general introduction to the PDG compilation. 
As above, our $N$ data points consist of $m=1,M$ distinct measured quantities,
each of which is measured by $n_m$ experiments, indexed by $k_m$.
($N = \sum_m n_m$).
Here it is useful to adopt the notation $x_{m{k_m}} \pm \sigma_{m{k_m}}$
for the individual measurements, and the notation $\bar{x}_m$ for 
the expected value for the $m^{\rm th}$ quantity.
After the fit is performed once, the error $\bar{\sigma}_m$ on $\bar{x}_m$
is evaluated from the output covariance matrix for the free fit parameters.
Then, the scale factor for the measured quantity $m$ is calculated as
\begin{equation}
S_m^2 = \frac{1}{n_m}\:\sum_{{k_m}=1}^{n_m}\:
\frac{(x_{m{k_m}}-\bar{x}_m)^2}{\sigma_{m{k_m}}^2 - \bar{\sigma}_m^2}.
\label{eq:scale}
\end{equation}
Next, the errors $\sigma_{m{k_m}}$ are scaled by the greater value
of $S_m$ and unity. For subsets of
correlated measurements (all from the same experiment), the index $k_m$
can be omitted to write $V_{mm'} = \rho_{mm'}\sigma_m\sigma_{m'}$;
the scale factors are applied to $\sigma_m$ and $\sigma_{m'}$ and 
\vec{V} and its inverse are recalculated.
Finally, the fit is performed a second time.
For each of the fit parameters, we report the central value from the 
first fit, and the error (and correlations) from the second fit. 
The scale factors for 
the errors on the fit parameters are defined as the ratios of the errors
from the second fit to those from the first. The value of $\chi^2$ reported 
is from the first fit. 

For the purposes of comparison, pull values are calculated for each
measurement simply as $(x_i - \bar{x}_i)/\sigma_i$.

For the BR/lifetime fits, the errors are in general symmetric 
to within rounding error; in any case we report the greater of the 
positive and negative \textsc{minos} errors.

\section{Fit for \boldmath{$K_L$} BRs and lifetime}
\label{app:BRLfit}
The 8 free parameters in the $K_L$ fit are
\BR{K_{e3}},  \BR{K_{\mu3}},  \BR{3\pi^0},  \BR{\pi^+\pi^-\pi^0},  
\BR{\pi^+\pi^-}, \BR{\pi^0\pi^0}, \BR{\gamma\gamma}, and $\tau_{K_L}$.
The fit makes use of the 18 measurements in \Tab{tab:KLdata}.
With one constraint, the fit has 11 degrees of freedom. 

The differences between our fit and the 2006 PDG fit are as follows:
\begin{itemize}
\item In our fit, the intermediate KTeV and KLOE values 
(i.e., before applying constraints) are the inputs, and the complete 
error matrix is used to handle the correlations between the measurements 
from each experiment.
In the 2006 PDG fit, the final KTeV and KLOE BR results were used
and one measurement involving \BR{3\pi^0} was removed in each case.
\item Our fit makes use of 
the preliminary \BR{3\pi^0} \cite{Lit04:ICHEP} 
and new \BR{\pi^+\pi^-}/\BR{K_{e3}} \cite{NA48+06:KLpp} measurements 
from NA48.
\item Our fit parameter \BR{\pi^+\pi^-} is understood to be inclusive
of the DE component. This helps to satisfy the constraint. The input data 
are treated consistently in this respect.
\item We do not make use of the measurement of 
\BR{\gamma\gamma}/\BR{\pi^0\pi^0}
from NA31 (Burkhardt '87), since both we and the PDG have excluded the 
other measurements from NA31.
\end{itemize}  

\begin{table}
\center
\begin{tabular}{clclc}
\hline\hline
Point & Parameter & Value & Source & Note \\[0.5ex]\hline
1  & $\tau_{K_L}$ & 50.92(30) ns & KLOE '05 & 1 \cite{KLOE+05:tauL}\\
2  & $\tau_{K_L}$ & 51.54(44) ns & Vosburgh '72 & \cite{Vosburg}\\
3  & $\BR{K_{e3}}$ & 0.4049(21) & KLOE '06 & 2 \cite{KLOE+06:BR}\\
4  & $\BR{K_{\mu3}}$ & 0.2726(16) & KLOE '06 & 2 \cite{KLOE+06:BR}\\
5  & $\BR{K_{\mu3}}/\BR{K_{e3}}$ & 0.6640(26) & KTeV '04 & 3 \cite{KTeV+04:BR}\\
6  & $\BR{3\pi^0}$ & 0.2018(24) & KLOE '06 & 2 \cite{KLOE+06:BR}\\
7  & $\BR{3\pi^0}/\tau_{K_L}$ & 3.795(58) MHz & NA48 '04 & 4 \cite{Lit04:ICHEP}\\
8  & $\BR{3\pi^0}/\BR{K_{e3}}$ & 0.4782(55) & KTeV '04 & 3 \cite{KTeV+04:BR}\\
9  & $\BR{\pi^+\pi^-\pi^0}$ & 0.1276(15) & KLOE '06 & 2 \cite{KLOE+06:BR}\\
10 & $\BR{\pi^+\pi^-\pi^0}/\BR{K_{e3}}$ & 0.3078(18) & KTeV '04 & 3 \cite{KTeV+04:BR}\\
11 & $\BR{\pi^+\pi^-}/\BR{K_{e3}}$ & 0.004856(29) & KTeV '04 & 3,5 \cite{KTeV+04:BR}\\ 
12 & $\BR{\pi^+\pi^-}/\BR{K_{e3}}$ & 0.004826(27) & NA48 '06 & 5 \cite{NA48+06:KLpp}\\
13 & $\BR{\pi^+\pi^-}/\BR{K_{\mu3}}$ & 0.007275(68) & KLOE '06 & 5 \cite{KLOE+06:KLpp}\\
14 & $\BR{K_{e3}}/\BR{\mbox{2 tracks}}$ & 0.4978(35) & NA48 '04 & 6 \cite{NA48+04:BR}\\
15 & $\BR{\pi^0\pi^0}/\BR{3\pi^0}$ & 0.004446(25) & KTeV '04 & 3 \cite{KTeV+04:BR}\\
16 & $\BR{\pi^0\pi^0}/\BR{\pi^+\pi^-} $ & 0.4391(13) & PDG '06 & 7 \cite{bib:pdg}\\
17 & $\BR{\gamma\gamma}/\BR{3\pi^0}$ & 0.00279(3) & KLOE '03 & \cite{KLOE03:gaga} \\
18 & $\BR{\gamma\gamma}/\BR{3\pi^0}$ & 0.00281(2) & NA48 '03 & \cite{NA4803:gaga}\\
\hline\hline
\end{tabular}
\caption{Input data used for the fit to $K_L$ BRs and lifetime.}
\label{tab:KLdata}
\end{table}

\subsubsection*{Notes on data in \Tab{tab:KLdata}}

\begin{enumerate}

\item Direct measurement of $\tau_{K_L}$ from $3\pi^0$ events; independent
      of other KLOE measurements \cite{KLOE+05:tauL}.

\item We make use of the KLOE results for the main $K_L$ BRs
      (\#3, \#4, \#6, and \#9) obtained before applying constraints
      on the sum of the BRs \cite{KLOE+06:BR}. 
      The BR values in \Tab{tab:KLdata} thus depend on $\tau_{K_L}$ as follows:
\begin{displaymath}
{\rm BR} = \frac{{\rm BR}^0}
                {1 + 0.0128~{\rm ns}^{-1}\:(\tau^0_{K_L} - \tau_{K_L})},
\end{displaymath}
      where $\tau^0_{K_L} = 51.54$~ns.
      The errors listed for these values in \Ref{KLOE+06:BR} include
      an explicit contribution from the uncertainty on the reference
      value of $\tau_{K_L}$. This contribution has been unfolded from
      the errors in \Tab{tab:KLdata}.
      In addition, these four BR measurements are correlated by common 
      systematics as described in KLOE Note 204, although the full
      correlation matrix is not given therein. The correlation matrix 
      is as follows:
\begin{displaymath}
\begin{array}{ccccc}
   & 3     & 4     & 6     & 9     \\
3  & 1.000 & 0.091 & 0.069 & 0.494 \\
4  &       & 1.000 &-0.025 & 0.267 \\
6  &       &       & 1.000 & 0.074 \\
9  &       &       &       & 1.000 \\
\end{array}
\end{displaymath}

\item The correlation matrix for the KTeV relative BR measurements
      (\#5, \#8, \#10, \#11, and \#15) is as follows \cite{KTeV+04:BR}:
\begin{displaymath}
\begin{array}{cccccc}
   & 5    & 8    & 10   & 11   & 15   \\
5  & 1.00 & 0.14 & 0.21 & 0.24 & 0.09 \\
8  &      & 1.00 &-0.06 &-0.07 & 0.30 \\
10 &      &      & 1.00 & 0.49 & 0.04 \\
11 &      &      &      & 1.00 & 0.07 \\
15 &      &      &      &      & 1.00 \\
\end{array}
\end{displaymath}

\item This is based on the preliminary NA48 measurement
      $\BR{3\pi^0} = 0.1966(34)$, as reported in \Ref{Lit04:ICHEP}.
      R.~Wanke has confirmed that the 2004 PDG value for $\tau_{K_L}$
      was used to obtain this result. The NA48 value for this BR
      scales directly with the lifetime value used. R.~Wanke has
      supplied the value in the table for the partial width, with
      the contribution to the error on the BR from the uncertainty
      on the $K_L$ lifetime unfolded. 

\item The fit value of \BR{\pi^+\pi^-} includes the DE component.
      The KLOE measurement of \BR{\pi^+\pi^-}/\BR{K_{\mu3}} 
      \cite{KLOE+06:KLpp} (\#13) is inclusive of DE. 
      The KTeV and NA48 measurements of \BR{\pi^+\pi^-}/\BR{K_{e3}}
      (\#11 and \#12, respectively) are treated as follows.
      \begin{itemize}
\item We use the 
      average values of ${\rm DE/(DE+IB)}$ from 
      \Refs{E731+93:ppg,KTeV+01:ppg} \andRef{KTeV+06:ppg}
      together with 
      $\BR{\pi^+\pi^-\gamma_{\rm IB}, E_\gamma>20~{\rm MeV}}/\BR{\pi^+\pi^-} =
      \SN{7.00}{-3}$ \cite{DAMS92:Hand}, to calculate that 
      DE accounts for $1.52(7)\%$ of the inclusive $K_L\to\pi^+\pi^-$ width.
      The error on this correction is negligible for the purposes of 
      the fit.
\item The KTeV measurement of 
      \BR{\pi^+\pi^-}/\BR{K_{e3}} (\#11) excludes DE (in the sense 
      that \Ref{KTeV+04:BR} says that DE is not in the generator for
      the acceptance calculation). We therefore subtract 1.52\%
      from the fit value of the ratio when calculating the 
      contribution to $\chi^2$ from this KTeV measurement.
\item The contribution from DE to the NA48 measurement
      $\BR{\pi^+\pi^-}/\BR{K_{e3}} = \SN{4.835(22)(20)}{-3}$
      is estimated to be $0.19(1)\%$, which we subtract to obtain value \#12.
      We then handle the data point in the same way as we do for KTeV.
      \end{itemize}

\item For our fit, $\BR{\mbox{2 tracks}}$ has to be calculated 
      from the free fit parameters. Like the PDG, we use
\begin{eqnarray*}
\BR{\mbox{2 tracks}}& =& \BR{K_{e3}} + \BR{K_{\mu3}} + 0.03508\,\BR{3\pi^0} \\
& & \mbox{}            + \BR{\pi^+\pi^-\pi^0} + \BR{\pi^+\pi^-}.
\end{eqnarray*}

\item From the ETAFIT analysis\cite{bib:pdg}.

\end{enumerate}

\subsection{Results}

The results of the fit are summarized in \Tab{tab:KLfit}.
The output correlation matrix is given in \Tab{tab:KLcorr}.
The pull values for the input measurements are listed in \Tab{tab:KLpull}.
With respect to the 2006 PDG fit, our
fit has a somewhat lower $\chi^2$ probability.
\begin{table}
\center
\begin{tabular}{lcccc}
\hline\hline
& \multicolumn{2}{c}{This fit}&\multicolumn{2}{c}{2006 PDG}\\
& \multicolumn{2}{c}{18 measurements}&\multicolumn{2}{c}{17 measurements} \\
& \multicolumn{2}{c}{$\chi^2/{\rm ndf} = 19.7/11$ (4.9\%)}&
  \multicolumn{2}{c}{$\chi^2/{\rm ndf} = 14.8/10$ (14.0\%)}\\\hline
Parameter            & Result               & $S$ & Result            & $S$ \\
\hline
\BR{K_{e3}}          & 0.4058(9)          & 1.3 & 0.4053(15)        & 2.1 \\
\BR{K_{\mu3}}        & 0.2706(8)          & 1.3 & 0.2702(7)         & 1.0 \\
\BR{3\pi^0}          & 0.1943(10)          & 1.3 & 0.1956(14)        & 1.9 \\
\BR{\pi^+\pi^-\pi^0} & 0.1259(8)          & 1.5 & 0.1256(5)         & 1.0 \\
\BR{\pi^+\pi^-}      & \SN{1.986(7)}{-3}  & 1.2 & \SN{1.976(8)}{-3} & 1.0 \\
\BR{\pi^0\pi^0}      & \SN{8.60(5)}{-4}   & 1.5 & \SN{8.69(4)}{-4}  & 1.1 \\
\BR{\gamma\gamma}    & \SN{5.45(4)}{-4}   & 1.1 & \SN{5.48(5)}{-4}  & 1.2 \\
\BR{\tau_{K_L}}      & 51.15(20)~ns       & 1.1 & 51.14(21)~ns      & 1.0 \\
\hline\hline
\end{tabular}
\caption{
Results of fit to $K_L$ BRs and lifetime, with comparison to 2006 PDG fit.}
\label{tab:KLfit}
\end{table}
\begin{table}
\center
\begin{tabular}{cccccccc}
$+1.000$ & $-0.286$ & $-0.422$ & $-0.288$ & $+0.112$ & $-0.282$ & $-0.270$ & $-0.005$\\
 & $+1.000$ & $-0.378$ & $-0.217$ & $-0.046$ & $-0.216$ & $-0.241$ & $+0.183$\\
 &  & $+1.000$ & $-0.354$ & $-0.029$ & $+0.609$ & $+0.637$ & $-0.036$\\
 &  &  & $+1.000$ & $-0.035$ & $-0.205$ & $-0.226$ & $-0.127$\\
 &  &  &  & $+1.000$ & $+0.205$ & $-0.020$ & $-0.033$\\
 &  &  &  &  & $+1.000$ & $+0.387$ & $-0.029$\\
 &  &  &  &  &  & $+1.000$ & $-0.027$\\
 &  &  &  &  &  &  & $+1.000$\\
\end{tabular}
\caption{Correlation matrix for output parameters of $K_L$ fit.}
\label{tab:KLcorr}
\end{table}
\begin{table}
\center
\begin{tabular}{cllc}
\hline\hline
Point & Parameter & Source & Pull \\[0.5ex]\hline
6  & $\BR{3\pi^0}$                       & KLOE '06     & $+2.74$\\
2  & $\tau_{K_L}$                        & Vosburgh '72 & $+0.88$\\
15 & $\BR{\pi^0\pi^0}/\BR{3\pi^0}$       & KTeV '04     & $+0.81$\\
9  & $\BR{\pi^+\pi^-\pi^0}$              & KLOE '06     & $+0.71$\\
4  & $\BR{K_{\mu3}}$                     & KLOE '06     & $+0.41$\\
18 & $\BR{\gamma\gamma}/\BR{3\pi^0}$     & NA48 '03     & $+0.31$\\
12 & $\BR{\pi^+\pi^-}/\BR{K_{e3}}$       & NA48 '06     & $+0.22$\\
7  & $\BR{3\pi^0}/\tau{K_L}$             & NA48 '04     & $-0.07$\\
8  & $\BR{3\pi^0}/\BR{K_{e3}}$           & KTeV '04     & $-0.13$\\
17 & $\BR{\gamma\gamma}/\BR{3\pi^0}$     & KLOE '03     & $-0.46$\\
16 & $\BR{\pi^0\pi^0}/\BR{\pi^+\pi^-}$   & PDG '06      & $-0.57$\\
14 & $\BR{K_{e3}}/\BR{\mbox{2 tracks}}$  & NA48 '04     & $-0.71$\\
1  & $\tau_{K_L}$                        & KLOE '05     & $-0.78$\\
13 & $\BR{\pi^+\pi^-}/\BR{K_{\mu3}}$     & KLOE '06     & $-0.94$\\
5  & $\BR{K_{\mu3}}/\BR{K_{e3}}$         & KTeV '04     & $-1.11$\\
11 & $\BR{\pi^+\pi^-}/\BR{K_{e3}}$       & KTeV '04     & $-1.32$\\
3  & $\BR{K_{e3}}$                       & KLOE '06     & $-1.37$\\
10 & $\BR{\pi^+\pi^-\pi^0}/\BR{K_{e3}}$  & KTeV '04     & $-1.39$\\
\hline\hline
\end{tabular}
\caption{Pull values for input data used in fit to $K_L$ BRs and lifetime.}
\label{tab:KLpull}
\end{table}

When our fit is run without inclusion of points \#7 and \#12,
without DE corrections for the $\pi^+\pi^-$ channel, and with
the measurement of \BR{\gamma\gamma}/\BR{\pi^0\pi^0} from NA31,
we reproduce the 2006 PDG fit result. 
In this configuration, the only difference between
our fit and the 2006 PDG fit is the treatment of the
BR and lifetime data from KLOE and KTeV. We obtain the
same values for all 8 fit parameters, with $\chi^2/{\rm ndf} = 14.9/10$.
Our scale factors in this case are more uniform than those 
obtained in the 2006 PDG fit; in particular, for 
\BR{K_{e3}}, \BR{3\pi^0}, and \BR{\pi^+\pi^-\pi^0} we have 
$S=1.2$, 1.1, and 1.4, to be compared with the second column of 
\Tab{tab:KLfit}.

Excluding the measurement of \BR{\gamma\gamma}/\BR{\pi^0\pi^0} from NA31
has a negligible effect on the fit results, while the number of degrees of 
freedom is reduced by one, giving $\chi^2/{\rm ndf} = 14.9/9$ (9.4\%).
Turning on the DE correction degrades the fit quality from 
$\chi^2/{\rm ndf} = 14.9/9$ to $19.6/9$ (2.02\%).
When points \#7 and \#12 are added, the fit quality is slightly improved
and the result in \Tab{tab:KLfit} is obtained.

However, the fit quality improves dramatically when the PDG ETAFIT
result for \BR{\pi^0\pi^0}/\BR{\pi^+\pi^-} (\#16) is removed. This is true
independently of whether or not the DE correction and/or the 
additional results are included.
For example, our same fit with the PDG ETAFIT point removed gives
$\chi^2/{\rm ndf} = 14.8/10$ (13.9\%), with changes in the fit values for 
the BRs at the $1\sigma$ level. In all other configurations (DE correction
on/off; points \#7, \#12, NA31 \BR{\gamma\gamma}/\BR{\pi^0\pi^0} 
included/excluded), the fit gives similar results.

Using the values of \BR{\pi^+\pi^-} and \BR{\pi^0\pi^0} from our 
fits including and excluding the PDG ETAFIT point, we have evaluated
\Reps\ from
\begin{displaymath}
\Reps = \frac{1}{6}\left[1 - R_S\:\frac{\BR{\pi^0\pi^0}}{\BR{\pi^+\pi^-}}\right]
\end{displaymath}
with $R_S\equiv\BR{K_S\to\pi^+\pi^-}/\BR{K_S\to\pi^0\pi^0}=2.2549(54)$
\cite{KLOE+06:KSpp} as described in \Sec{sec:KS}. We obtain
\begin{displaymath}
\begin{array}{lcll}
\Reps & = & \SN{(-25\pm23)}{-4} & \hspace{1cm}\mbox{(without ETAFIT);}\\
\Reps & = & \SN{(14\pm11)}{-4} & \hspace{1cm}\mbox{(with ETAFIT);}\\
\end{array}
\end{displaymath}
to be compared to the current PDG average, \SN{(16.7\pm2.3)}{-4}.  
The ETAFIT point is very precise; when it is included, the fit results 
for \BR{\pi^0\pi^0}/\BR{\pi^+\pi^-} are highly constrained. This 
pulls down the value of \BR{\pi^0\pi^0}, and, also of \BR{3\pi^0}, 
via the KTeV measurement of \BR{\pi^0\pi^0}/\BR{3\pi^0}.
As seen from \Tab{tab:KLpull}, the measurement with the largest 
positive pull on the fit is the KLOE measurement of \BR{3\pi^0}, which
PDG has chosen to eliminate from the 2006 fit as part of their treatment 
of the correlated KLOE measurements.  

We emphasize that the values of \BR{K_{e3}} and \BR{K_{\mu3}} are
not affected very much by these developments. The scale factor, and
hence the reported error, on \BR{K_{e3}} is significantly smaller in
our fit, which spreads the pulls somewhat more evenly over the different
measurements than does the PDG fit.

\section{Fit for \boldmath{$K^\pm$} BRs and lifetime}
\label{sec:KpmBR}

The 7 free parameters in the $K^\pm$ fit are
\BR{K_{\mu2}}, \BR{\pi\pi^0},  \BR{\pi\pi\pi},  \BR{K_{e3}}, 
\BR{K_{\mu3}}, \BR{\pi\pi^0\pi^0}, and $\tau_{K^\pm}$.
The fit makes use of the 26 measurements in \Tab{tab:Kpmdata}.
With one constraint, the fit has 24 degrees of freedom. 
\begin{table}
\center
\begin{tabular}{clclc}
\hline\hline
Point & Parameter & Value & Source & Note \\[0.5ex]\hline
1  & $\tau_{K^\pm}$ & 12.451(30) ns & Koptev '95   \cite{Koptev:1995}           & 1   \\
2  & $\tau_{K^\pm}$ & 12.368(41) ns & Koptev '95   \cite{Koptev:1995}           & 1   \\
3  & $\tau_{K^\pm}$ & 12.380(16) ns & Ott '71       \cite{Ott:1971}             &     \\
4  & $\tau_{K^\pm}$ & 12.272(36) ns & Lobkowicz '69 \cite{Lobkowicz:1970}       &     \\
5  & $\tau_{K^\pm}$ & 12.443(38) ns & Fitch '65    \cite{Fitch:1965}            &     \\
6  & $\tau_{K^\pm}$ & 12.367(78) ns & KLOE '06      \cite{KLOE:taupm}           & 2   \\
7  & $\tau_{K^\pm}$ & 12.391(55) ns & KLOE '07    \cite{KLOE:taupm}             & 2   \\
8  & $\BR{K_{\mu2}}$ & 0.6366(17) & KLOE '06      \cite{KLOE05:kl2}             &     \\
9  & $\BR{\pi\pi^0}$ & 0.2066(11) & KLOE '07      \cite{KLOE:pipo}              &     \\
10 & $\BR{\pi\pi^0}/\BR{K_{\mu2}}$ & 0.3329(48) & Usher '92 \cite{Usher:1992}   &     \\
11 & $\BR{\pi\pi^0}/\BR{K_{\mu2}}$ & 0.3355(57) & Weissenberg '76\cite{Weissenberg:1976} &     \\
12 & $\BR{\pi\pi^0}/\BR{K_{\mu2}}$ & 0.3277(65) & Auerbach '67 \cite{Auerbach:1967} &     \\
13 & $\Gamma(\pi\pi\pi)$ & 4.513(24)~MHz & Ford '70    \cite{Ford:1970}            &     \\
14 & $\BR{K_{e3}}$ & 0.04965(53) & KLOE '07           \cite{KLOE:kl3pm}            & 2,4 \\
15 & $\BR{K_{e3}}/\BR{\pi\pi^0\!+K_{\mu3}+\!\pi2\pi^0}$ & 0.1962(36) & Sher '03 \cite{Sher:2003} &     \\
16 & $\BR{K_{e3}}/\BR{K_{\mu2}+\pi\pi^0}$ & 0.0616(22) & Eschstruth '68\cite{Eschtruth:1968} &     \\
17 & $\BR{K_{e3}}/\BR{K_{\mu2}+\pi\pi^0}$ & 0.0589(21) & Cester '66\cite{Cester:1966}  &     \\
18 & $\BR{K_{e3}}/\BR{\pi\pi^0}$ & 0.2449(16) & ISTRA '07    \cite{Rom06:ke3}   & 2   \\
19 & $\BR{K_{e3}}/\BR{\pi\pi^0}$ & 0.2470(10) & NA48 '07    \cite{NA48+07:BR}   & 5   \\
20 & $\BR{K_{\mu3}}$ & 0.03233(39) & KLOE '07             \cite{KLOE:kl3pm}     & 2,4 \\
21 & $\BR{K_{\mu3}}/\BR{\pi\pi^0}$ & 0.1636(7) & NA48 '07 \cite{NA48+07:BR}     & 5   \\
22 & $\BR{K_{\mu3}}/\BR{K_{e3}}$ & 0.671(11) & Horie '01  \cite{Horie:2001}     &     \\
23 & $\BR{K_{\mu3}}/\BR{K_{e3}}$ & 0.670(14) & Heintze '77 \cite{Heintze:1977}   &     \\
24 & $\BR{K_{\mu3}}/\BR{K_{e3}}$ & 0.667(17) & Botterill '68 \cite{Botteril:1968}&     \\
25 & $\BR{\pi\pi^0\pi^0}$ & 0.01763(26) & KLOE '04     \cite{KLOE04:taup}        &     \\
26 & $\BR{\pi\pi^0\pi^0}/\BR{\pi\pi\pi}$ & 0.303(9) & Bisi '65  \cite{Bisi:1965} &     \\
\hline\hline
\end{tabular}
\caption{Input data used for the fit to $K^\pm$ BRs and lifetime.}
\label{tab:Kpmdata}
\end{table}

The principal difference between the fit 
performed here and the 2006 PDG fit is that our fit includes the 
following recent measurements:
\begin{itemize}
\item Preliminary $\tau_{K^\pm}$ from KLOE (\#6, \#7);
\item Preliminary \BR{K_{e3}} and \BR{K_{\mu3}} from KLOE (\#14, \#20);
\item Preliminary \BR{\pi\pi^0} from KLOE (\#9);
\item Preliminary \BR{K_{e3}}/\BR{\pi\pi^0} from ISTRA+ (\#18);
\item \BR{K_{e3}}/\BR{\pi\pi^0} (\#19) and \BR{K_{\mu3}}/\BR{\pi\pi^0} 
from NA48/2 (\#21).
\end{itemize}
These new measurements have a profound impact on the results of the fit.
Other differences are as follows.
\begin{itemize}
\item In the 2006 PDG fit, \BR{\pi^0\pi^0e\nu} is a free parameter (but 
curiously, \BR{\pi\pi e\nu}, for which there is a published measurement
from E865 with much higher accuracy \cite{E865+03:Ke4}, is not). The
PDG fit therefore uses three measurements involving
\BR{\pi^0\pi^0e\nu} and \BR{\pi^0\pi^0e\nu}/\BR{K_{e3}} that are not
used in our fit.
\item We don't use the six BR measurements from Chiang '72.
Our reading of Chiang '72 suggests that no attempt was made
      to implement radiative corrections for the branching ratio 
      analysis. In addition, the six BR measurements from Chiang '72
      are constrained to sum to unity. The correlation matrix 
      is not available. PDG omits \BR{\pi\pi\pi}.

\end{itemize}
It would be highly desirable to discard many other old measurements in the 
$K^\pm$ fit as 2006 PDG has done for the $K_L$ fit.
Unfortunately,  are no recent measurements
involving \BR{\pi\pi\pi}. As a result, the fit is unstable if only 
recent measurements are used.

\subsubsection*{Notes on data in \Tab{tab:Kpmdata}}

\begin{enumerate}

\item The only difference between the Koptev measurements is the 
      material used for the kaon stopper (\#1--U, \#2--Cu).

\item Preliminary measurement.

\item The dependence of these BRs on the $K^\pm$ lifetime is accounted 
      for in the fit:
\begin{displaymath}
{\rm BR} = {\rm BR}^0\:[1 + 0.0405(\tau_{K^\pm} - \tau_{K^\pm}^0)]
\end{displaymath}
      where ${\rm BR}^0$ is evaluated with $\tau_{K^\pm} = 12.360$~ns.
      The uncertainty from the value of 
      $\tau_{K^\pm}$ may not have been properly unfolded.
      In addition, these two 
      measurements are have a correlation coefficient of 0.627, 
      mainly from the use of common efficiency corrections.

\item The recent NA48 publication \cite{NA48+07:BR} gives values for
      \BR{K_{e3}}/\BR{\pi\pi^0},\\ \BR{K_{\mu3}}/\BR{\pi\pi^0}.
      The value of \BR{K_{e3}}/\BR{\pi\pi^0} has been updated at KAON07.

\end{enumerate}

\subsection{Results}
The results of the fit are summarized in \Tab{tab:Kpmfit}.
The output correlation matrix is given in \Tab{tab:Kpmcorr}.
The pull values for the input measurements are listed in \Tab{tab:Kpmpull}.
The poor fit quality derives from the following sources.
\begin{itemize}
\item The fit quality is significantly degraded by the 
scatter in the five older measurements of $\tau_{K^\pm}$; when these are 
replaced with their PDG average with scaled error, 
$\tau_{K^\pm} = 12.385(25)$~ns,
the fit gives $\chi^2/{\rm ndf} = 24.3/16$ (8.4\%), with no significant
changes in the results. Note that after this treatment the fit quality is
about the same as it is for the 2006 PDG fit (which, however,
includes all of the older 
$\tau_{K^\pm}$ measurements without taking the average).

\item There is some conflict among the newer measurements involving 
\BR{K_{e3}}, as seen from the pulls for the NA48 '07 (\#19), Sher '03 (\#15), 
ISTRA '07 (\#20), and KLOE '07 (\#14) measurements: $+1.04$, $-0.26$, $-0.74$,
and $-2.13$, respectively.(\Tab{tab:Kpmpull}).
\end{itemize}

The evolution of the average values of the BRs for 
$K^\pm_{\ell3}$ decays and for 
the important normalization channels as a result of the introduction
of the preliminary measurements is evident in \Fig{fig:kpmavg}. 
The figure dramatically illustrates why experiments that measure 
ratios such as $\BR{K_{e3}}/\BR{\pi\pi^0}$ should {\em always} quote
the ratio with usable errors, in addition to the normalized, final value.

\begin{table}
\center
\begin{tabular}{lcccc}
\hline\hline
& \multicolumn{2}{c}{This fit}&\multicolumn{2}{c}{2006 PDG}\\
& \multicolumn{2}{c}{26 measurements}&\multicolumn{2}{c}{26 measurements} \\
& \multicolumn{2}{c}{$\chi^2/{\rm ndf} = 42/20$ (0.31\%)}&
  \multicolumn{2}{c}{$\chi^2/{\rm ndf} = 30/19$ (5.2\%)}\\\hline
Parameter           & Result          & $S$ & Result        & $S$ \\
\hline
\BR{K_{\mu2}}       & 63.57(11)\%   & 1.1 & 63.44(14)\%   & 1.2 \\
\BR{\pi\pi^0}       & 20.64(8)\%   & 1.1 & 20.92(12)\%   & 1.1 \\
\BR{\pi\pi\pi}      & 5.595(31)\%   & 1.0 & 5.590(31)\%   & 1.1 \\
\BR{K_{e3}}         & 5.078(25)\%     & 1.2 & 4.98(7)\%     & 1.3 \\
\BR{K{\mu3}}        & 3.365(27)\%   & 1.7 & 3.32(6)\%     & 1.2 \\
\BR{\pi\pi^0\pi^0}  & 1.750(24)\%   & 1.1 & 1.757(24)\%   & 1.1 \\
\BR{\pi^0\pi^0e\nu} & \multicolumn{2}{c}{Not in fit}
                                      & \SN{2.2(4)}{-5}     & 1.0 \\
\BR{\tau_{K^\pm}}   & 12.384(19)~ns & 1.7 & 12.385(24)~ns & 2.1 \\
\hline\hline
\end{tabular}
\caption{
Results of fit to $K^\pm$ BRs and lifetime, with comparison to 2006 PDG fit.}
\label{tab:Kpmfit}
\end{table}

\begin{table}
\center
\begin{tabular}{ccccccc}
$ 1.000$ & $-0.874$ & $-0.170$ & $-0.725$ & $-0.548$ & $-0.258$ & $-0.045$\\
 & $ 1.000$ & $-0.121$ & $ 0.610$ & $ 0.333$ & $ 0.031$ & $-0.032$\\
 &  & $ 1.000$ & $-0.100$ & $-0.074$ & $ 0.055$ & $ 0.273$\\
 &  &  & $ 1.000$ & $ 0.442$ & $ 0.009$ & $-0.030$\\
 &  &  &  & $ 1.000$ & $-0.010$ & $-0.020$\\
 &  &  &  &  & $ 1.000$ & $ 0.010$\\
 &  &  &  &  &  & $ 1.000$\\
\end{tabular}
\caption{Correlation matrix for output parameters of $K^\pm$ fit.}
\label{tab:Kpmcorr}
\end{table}

\begin{table}
\center
\begin{tabular}{cllc}
\hline\hline
Point & Parameter & Source & Pull \\[0.5ex]\hline
1  & $\tau_{K^\pm}$                                      & Koptev '95      &$ +2.25$\\
11 & $\BR{\pi\pi^0}/\BR{K_{\mu2}}$                       & Weissenberg '76 &$ +1.89$\\
10 & $\BR{\pi\pi^0}/\BR{K_{\mu2}}$                       & Usher '92       &$ +1.70$\\
5  & $\tau_{K^\pm}$                                      & Fitch '65       &$ +1.56$\\
21 & $\BR{K_{\mu3}}/\BR{\pi\pi^0}$                       & NA48 '07        &$ +1.04$\\
19 & $\BR{K_{e3}}/\BR{\pi\pi^0}$                         & NA48 '07        &$ +1.03$\\
22 & $\BR{K_{\mu3}}/\BR{K_{e3}}$                         & Horie '01       &$ +0.76$\\
16 & $\BR{K_{e3}}/\BR{K_{\mu2}+\pi\pi^0}$                & Eschstruth '68  &$ +0.59$\\
8  & $\BR{K_{\mu2}}$                                     & KLOE '06        &$ +0.52$\\
23 & $\BR{K_{\mu3}}/\BR{K_{e3}}$                         & Heintze '77     &$ +0.52$\\
25 & $\BR{\pi\pi^0\pi^0}$                                & KLOE '04        &$ +0.52$\\
12 & $\BR{\pi\pi^0}/\BR{K_{\mu2}}$                       & Auerbach '67    &$ +0.46$\\
24 & $\BR{K_{\mu3}}/\BR{K_{e3}}$                         & Botterill '68   &$ +0.26$\\
7  & $\tau_{K^\pm}$                                      & KLOE '07        &$ +0.14$\\
13 & $\Gamma(\pi\pi\pi)$                                 & Ford '70        &$ -0.22$\\
6  & $\tau_{K^\pm}$                                      & KLOE '06        &$ -0.21$\\
3  & $\tau_{K^\pm}$                                      & Ott '71         &$ -0.22$\\
15 & $\BR{K_{e3}}/\BR{K_{\mu3}+\!\pi\pi^0+\!\pi2\pi^0}$  & Sher '03        &$ -0.26$\\
2  & $\tau_{K^\pm}$                                      & Koptev '95      &$ -0.38$\\
17 & $\BR{K_{e3}}/\BR{K_{\mu2}+\pi\pi^0}$                & Cester '66      &$ -0.67$\\
20 & $\BR{K_{e3}}/\BR{\pi\pi^0}$                         & ISTRA '07       &$ -0.74$\\
26 & $\BR{\pi\pi^0\pi^0}/\BR{\pi\pi\pi}$                 & Bisi '65        &$ -1.07$\\
14 & $\BR{K_{e3}}$                                       & KLOE '07        &$ -2.13$\\
4  & $\tau_{K^\pm}$                                      & Lobkowicz '71   &$ -3.10$\\
20 & $\BR{K_{\mu3}}$                                     & KLOE '07        &$ -3.41$\\
\hline\hline
\end{tabular}
\caption{Pull values for input data used in fit to $K^\pm$ BRs and lifetime.}
\label{tab:Kpmpull}
\end{table}

\section{Averages of form-factor slopes}
\label{app:ff}
\subsection{Procedure}

We work principally with quadratic form-factor slope parametrization.
To average the form-factor slopes, a $\chi^2$ fit
with correlations is performed.
Scale factors for the errors are calculated as described in section~\ref{app:fitprocedure}.
For the fit to the form-factor slopes, since there are no measurements
of combinations of the fit parameters, the scale factors can be obtained
directly from \Eq{eq:scale}.
Because of the high degree of correlation in the measurements of $\lambda'$
and $\lambda''$, a large scale factor may result in a small change in
$\chi^2$ from the fits. We therefore report scaled errors only when the
value of $\chi^2/{\rm ndf}$ is unsatisfactory.

\subsection{Input data}
The data used in the fit are summarized 
in \Tab{tab:FF}.
\begin{sidewaystable}[htb]
\center
\begin{tabular}{@{}lcccccccc@{}}
\hline\hline
Experiment & \SN{\lambda_+'}{3} & \SN{\lambda_+''}{3} & 
 \SN{\lambda_0}{3} &
  $\rho(\lambda_+',\lambda_+'')$ & $\rho(\lambda_+',\lambda_0)$ &
  $\rho(\lambda_+'',\lambda_0)$ & Analysis & Note \\
\hline
KLOE $K_L$ $e3$ \cite{KLOE+06:FF} &
     $25.5\pm1.8$ & $1.4\pm0.8$ & & $-0.95$ & & & $t$ from $K_S\to\pi^+\pi^-$ \\
KLOE $K_L$ $\mu3$ \cite{KLOE+07:m3FF} &
     $22.3\pm10.5$ & $4.8\pm5.2$ & $9.1\pm6.5$ &
     $-0.97$ & $+0.81$ & $-0.91$ & $E_\nu^*$ \\
KLOE $K_L$ $e3$-$\mu3$ \cite{KLOE+07:m3FF} &
     $25.6\pm1.8$ & $1.5\pm0.8$ & $15.4\pm2.2$ &
     $-0.95$ & $+0.29$ & $-0.38$ & average & 1 \\
KTeV $K_L$ $e3$ \cite{KTeV+04:FF} &
     $21.67\pm1.99$ & $2.87\pm0.78$ & &
     $-0.97$ & & & $t_\perp^\pi$ \\
KTeV $K_L$ $\mu3$ \cite{KTeV+04:FF} &
     $17.03\pm3.65$ & $4.43\pm1.49$ & $12.81\pm1.83$ &
     $-0.96$ & $+0.65$ & $-0.75$ & $(t_\perp^\mu,M_{\pi\mu})$ \\
KTeV $K_L$ $e3$-$\mu3$ \cite{KTeV+04:FF} &
     $20.64\pm1.75$ & $3.20\pm0.69$ & $13.72\pm1.31$ &
     $-0.97$ & $+0.34$ & $-0.44$ & average & 1 \\
NA48 $K_L$ $e3$ \cite{NA48+04:e3FF} &
     $28.0\pm2.4$ & $0.4\pm0.9$ & & $-0.88^*$ & & & $(E_\nu^*,t_{\rm low},t_{\rm high})$ & 2 \\
NA48 $K_L$ $\mu3$ \cite{NA48+06:m3FF} &
     $20.5\pm3.3$ & $2.6\pm1.3$ & $9.5\pm1.4$ &
     $-0.96$ & $+0.63$ & $-0.73$ & $(y,z)_{\rm low}$ \\
ISTRA+ $K^-$ $e3$ \cite{ISTRA+04:e3FF} &
     $24.85\pm1.66$ & $1.92\pm0.94$ & & $-0.95*$ & & & $(y,z)_{\rm 2C fit}$ & 3 \\
ISTRA+ $K^-$ $\mu3$ \cite{ISTRA+04:m3FF} &
     $22.99\pm6.42^*$ & $2.29\pm2.29^*$ & $17.11\pm2.25^*$ &
     $-0.82^*$ & $-0.12^*$ & $-0.41^*$ & $(y,z)_{\rm 2C fit}$ & 4 \\
\hline\hline
\end{tabular}
\caption{Measurements of $K_{\ell3}$ form-factor slopes. Values marked
with an asterisk involve additional assumptions; see notes in text.}
\label{tab:FF}
\end{sidewaystable}
The following notes apply to the table entries.

\begin{enumerate}
 \item In our combined fits to $K_{e3}$ and $K_{\mu3}$ data, we use the 
 averages 
 quoted by KLOE and KTeV rather than using their $K_{e3}$ and $K_{\mu3}$ 
 measurements
 separately. In any event, our averages of the $K_{e3}$ and $K_{\mu3}$ 
 results from 
 each experiment have good values of $\chi^2/{\rm ndf}$ and
 confirm the results quoted by the experiments, including the correlation
 coefficients.

 \item The exact value of $\rho(\lambda_+',\lambda_+'')$ is not available for
 the NA48 $K_{e3}$ measurement. NA48 and PDG together estimated $\rho=-0.88$;
 this value appears in the 2006 PDG listings \cite{bib:pdg}. 
 For use with \Eq{eq:Taylor},
 we put $\lambda_+'' = 2\lambda_+''^{\rm (NA48)}$.

 \item An official value of $\rho(\lambda_+',\lambda_+'')$ is not available for
 the ISTRA+ $K_{e3}$ measurement; the value in the Table was obtained directly
 from the collaboration. 
 For use with \Eq{eq:Taylor},
 we put $\lambda_+' = C\,\lambda_+'^{\rm (ISTRA)}$ and
 $\lambda_+'' = 2C^2\lambda_+''^{\rm (ISTRA)}$,
 with $C = (m_{\pi^+}/m_{\pi^0})^2 = 1.069223$.

 \item Systematic errors for the ISTRA+ quadratic fit results for $K_{\mu3}$
 are not given in
 \Ref{ISTRA+04:m3FF}; the errors in the table are statistical only.
 Nor are the correlation coefficients available; these have been
 obtained directly from the collaboration.
 For use with \Eq{eq:Taylor},
 $\lambda_+'$ and $\lambda_+''$ are converted as above; we also put
 $\lambda_0 = C\,\lambda_0^{\rm (ISTRA)}.$
 Finally, we note that no information concerning the treatment of
 radiative corrections
 is given in \Ref{ISTRA+04:m3FF}. Failure to account for radiative effects could
 result in a noticeable systematic shift in the slope results.
\end{enumerate}


For the KLOE and KTeV form-factor slope measurements, 
the correlation coefficients
apply to the total errors (statistical and systematic).
For the ISTRA+ and NA48 $K_{\mu3}$ slopes, the correlation 
coefficients appear to
apply to the statistical errors.
In our fits, we assume that the correlation coefficients apply
to the total errors on the form-factor slopes (statistical and systematic).
This approximation is motivated as follows.
In general, the systematic errors are
estimated by varying analysis parameters and refitting. In that case,
the statistical correlations naturally present will also affect the
excursions due to systematic variations, see Appendix~\ref{app:Fslope}.


\subsection{Fit results for \boldmath{$K_{\ell3}$} slopes excluding NA48 \boldmath{$K_{\mu3}$} data}
The result of our fit to all data is presented in \Tab{tab:l3ff}.
As discussed in \Sec{sec:l3ff}, the NA48 $K_{\mu3}$ form-factor slope measurements
are in contrast with the results from the other experiments. As an exercise,
we fit all results in \Tab{tab:FF} except the NA48 measurement
of the $K_{\mu3}$ slopes \cite{NA48+06:m3FF}.
The results are shown in \Tab{tab:oldKl3FF}. The first column of the
table gives the results of the fit to all other measurements from KLOE; 
the second gives the results of the fit to the
$K_L$ measurements from KLOE, KTeV, and the $K_{L\:e3}$ measurement from
NA48.
\TABLE{
\begin{tabular}{lccc}
\hline\hline
                                 & $K_L$ and $K^-$   & $K_L$ only     \\
\hline
Measurements                     & 13		     & 8		\\    	
$\chi^2/{\rm ndf}$               & 13/9 (24.9\%)     & 9/5 (12.3\%)   \\
$\lambda_+'\times 10^3 $         & $25.0\pm0.8$      & $24.5\pm1.1$ \\
$\lambda_+''\times 10^3 $        & $1.6\pm0.4$       & $1.8\pm0.4$  \\
$\lambda_0\times 10^3  $         & $16.0\pm0.8$      & $14.8\pm1.1$ \\
$\rho(\lambda_+',\lambda_+'')$   & $-0.94$           & $-0.95$         \\
$\rho(\lambda_+',\lambda_0)$     & $+0.26$           & $+0.28$         \\
$\rho(\lambda_+'',\lambda_0)$    & $-0.37$           & $-0.38$         \\
$I(K^0_{e3})$                    & 0.15459(20)       & 0.15446(27)    \\
$I(K^\pm_{e3})$                  & 0.15894(21)       & 0.15881(28)    \\
$I(K^0_{\mu3})$                  & 0.10268(20)       & 0.10236(28)    \\
$I(K^\pm_{\mu3})$                & 0.10559(20)       & 0.10532(29)    \\
$\rho(I_{e3},I_{\mu3})$          & $+0.59$          & $+0.62$         \\
\hline\hline
\end{tabular}
\caption{Averages of quadratic fit results for $K_{e3}$ and $K_{\mu3}$ slopes,
excluding new $K_{\mu3}$ data from NA48.}
\label{tab:oldKl3FF}}
The evaluations
of the phase-space integrals for all four modes are listed in each case.
Correlations are fully accounted for, both in the fits and in the evaluation
of the integrals.
The values of $\chi^2/{\rm ndf}$ do not raise any significant concerns
about the compatibility of the input data. The fit to all data gives
$\chi^2/{\rm ndf} = 12.6/10$ (25.0\%).

The evaluations
of the phase-space integrals for all four modes are listed in each case.
Correlations are fully accounted for, both in the fits and in the evaluation
of the integrals.
The values of $\chi2/{\rm ndf}$ do not raise any significant concerns
about the compatibility of the input data. The fit to all data gives
$\chi2/{\rm ndf} = 12/9$ (22.3\%).

{

\def \gev  {{\rm \,Ge\kern-0.125em V}}
\def \mev  {{\rm \,Me\kern-0.125em V}}
\def \kev  {{\rm \,ke\kern-0.125em V}}
\def \ev   {{\rm \,e\kern-0.125em V}}
\newcommand{\Dcalo}{\ensuremath{\Delta t_\mathrm{cal}}}
\newcommand{\dpstar}{\ensuremath{\Delta p^*}}
\newcommand{\dTCA}{\ensuremath{d_{\mathrm{TC}}}}
\newcommand{\dtTCA}{\ensuremath{d_{\perp,\,\mathrm{TC}}}}
\newcommand{\DTOF}{\ensuremath{\Delta\mathrm{TOF}}}
\newcommand{\efv}{\ensuremath{\varepsilon_{\mathrm{FV}}}}
\newcommand{\pComp}{\ensuremath{p_\mathrm{MC tag}^*}}
\newcommand{\pPCA}{\ensuremath{\mathbf{p}_\mathrm{PCA}}}
\newcommand{\pstarMC}{\ensuremath{p_\mathrm{MC}^*}}
\newcommand{\rDC}{\ensuremath{r_\mathrm{DC}}}
\newcommand{\rPCA}{\ensuremath{r_\mathrm{PCA}}}
\newcommand{\Tcl}{\ensuremath{t_\mathrm{cl}}}
\newcommand{\Ttrig}{\ensuremath{t_\mathrm{trig}}}
\newcommand{\trLen}{\ensuremath{L}}
\newcommand{\xPCA}{\ensuremath{\mathbf{x}_\mathrm{PCA}}}
\newcommand{\xc}{\ensuremath{\mathbf{x}_\mathrm{c}}}
\newcommand{\pc}{\ensuremath{\mathbf{p}_\mathrm{c}}}
\newcommand{\lc}{\ensuremath{l_\mathrm{c}}}
\newcommand{\dc}{\ensuremath{d_\mathrm{c}}}
\newcommand{\tof}{\mbox{TOF}}
\newcommand{\klpln}{\mbox{$K_{L}\to\pi^{\pm}\ell^{\mp}\nu$}}
\newcommand{\klpmn}{\mbox{$K_{L}\to\pi\mu\nu$}}
\newcommand{\semil}{\mbox{$K_{\mu3}$}}
\renewcommand{\vec}[1]{\mbox{\boldmath{$\rm#1$}}}
\newcommand{\unit}[1]{\mbox{\boldmath{$\rm\hat{#1}$}}}
\newcommand{\fphat}{\mbox{$\tilde{f}_+(t)$}}
\newcommand{\fpha}{\mbox{$\tilde{f}_+$}}

\newcommand{\fzhat}{\mbox{$\tilde{f}_0(t)$}}
\newcommand{\fmhat}{\mbox{$\tilde{f}_-(t)$}}

\renewcommand{\textfraction}{0.1}
\renewcommand{\topfraction}{0.9}
\renewcommand{\bottomfraction}{0.8}
\renewcommand{\floatpagefraction}{0.75}

\newcommand{\pemiss}{\ensuremath{|{\bf p}_{\rm miss}|-E_{\rm miss}}}
\newcommand{\dmp}{\ensuremath{\Delta_{\mu \pi}}}
\newcommand{\dep}{\ensuremath{\Delta_{e \pi}}}
\newcommand{\pmiss}{\ensuremath{{\bf p}_{\rm miss}}}
\newcommand{\emiss}{\ensuremath{E_{\rm miss}}}
\newcommand{\LK}{\ensuremath{L_{K}}}
\newcommand{\miss}{\mbox{$E_{\rm miss}-p_{\rm miss}$}}

\newcommand{\kspp}{\mbox{$K_{S}\to\pi^+\pi^- $}}
\newcommand{\klppp}{\mbox{$K_{L}\to\pi^+\pi^-\pi^0$}}
\newcommand{\klpp}{\mbox{$K_{L}\to\pi^+\pi^- $}}
\newcommand{\klpen}{\mbox{$K_{L}\to\pi^{\pm}e^{\mp}\nu$}}
\newcommand{\aff}[2]{Dipartimento di Fisica dell'Universit\`a #1 e Sezione INFN, #2, Italy.}
\newcommand{\affd}[1]{Dipartimento di Fisica dell'Universit\`a e Sezione INFN, #1, Italy.}

\font\euler=eufm10 at 12pt
\def\Ma{\hbox{\euler M}}
\def\ifm#1{\relax\ifmmode#1\else$#1$\fi} \def\DAF{DA\char8NE} \def\x{\ifm{\times}}
\def\pt#1,#2,{\ifm{#1\x10^{#2}}} \def\up#1{\ifm{^{#1}}} \def\dn#1{\ifm{_{#1}}}
\def\ab{\ifm{\sim}} \def\deg{\ifm{^\circ}} \def\gam{\ifm{\gamma}} \def\to{\ifm{\rightarrow}}
\def\kl{\ifm{K_L}} \def\ks{\ifm{K_S}} \def\eiii{\ifm{\pi^\pm e^\mp\nu}} \def\keiii{\ifm{K_{e3}}}
\def\muiii{\ifm{\pi^\pm \mu^\mp\nu}} \def\kmuiii{\ifm{K_{\mu3}}} \def\pio{\ifm{\pi^0\pi^0}}
\def\po{\ifm{\pi^0}} \def\pic{\ifm{\pi^+\pi^-}} \def\K{\ifm{K}}   \def\ff{$\phi$--factory}
\def\rmk{\rm\kern.5mm }   \def\dif{\hbox{d}} \def\f{\ifm{\phi}}  \def\minus{$-$}
\def\Vus{\ifm{|V_{us}|}} \def\bye{\end{document}}

\catcode`@=11 
\newdimen\z@ \z@=0pt 
\newskip\z@skip \z@skip=0pt plus0pt minus0pt
\def\m@th{\mathsurround=\z@}
\def\ialign{\everycr{}\tabskip\z@skip\halign} 
\def\eqalign#1{\null\,\vcenter{\openup\jot\m@th
  \ialign{\strut\hfil$\displaystyle{##}$&$\displaystyle{{}##}$\hfil
      \crcr#1\crcr}}\,}
\catcode`@=12 
\def\figb#1;#2;{\parbox{#2cm}{\epsfig{file=#1.eps,width=#2cm}}}
\let\cl=\centerline
\def\figbc#1;#2;{\cl{\figb #1;#2;}}

\newcommand{\lam}{\mbox{$\lambda_+$}}
\newcommand{\lamp}{\mbox{$\lambda'_+$}}
\newcommand{\lampp}{\mbox{$\lambda''_+$}}
\newcommand{\lamo}{\mbox{$\lambda_0$}}
\newcommand{\lamop}{\mbox{$\lambda'_0$}}
\newcommand{\lamopp}{\mbox{$\lambda''_0$}}
\newcommand{\jth}{\ensuremath{j^\mathrm{th}}}
\def\dlp{\hbox{$\delta\lambda_+'$}}
\def\dlpp{\hbox{$\delta\lambda_+''$}} \def\dop{\hbox{$\delta\lambda_0'$}}
\def\dopp{\hbox{$\delta\lambda_0''$}}
\def\eqq#1{(\ref{#1})}
\def\eq#1{(eq.~\ref{#1})}

\section{Error estimates}
\label{app:Fslope}
It is quite easy to estimate the ideal error in the measurements of a set of parameters {\bf p}=$(p_1,\:p_2,\ldots\:p_n)$ from fitting some distribution function to experimentally determined spectra.
Let $F(\mathbf{p},x)$ be a probability density function, PDF, where ${\mathbf p}$ is some parameter vector, which we want to determine and $x$ is a running variable, like $t$. The inverse of the covariance matrix for the maximum likelihood estimate of the parameters is given by \cite{hcra}:
$$({\bf G}^{-1})_{ij}=-{\partial^2 \ln L\over\partial p_i\partial p_j}$$
from which, for $N$ events, it trivially follows:
$$\left({\bf G}^{-1}\right)_{ij}=N\:\int{1\over F}\,{\partial F\over\partial p_i }\,{\partial F\over\partial p_j}\:\dif\upsilon,$$
with $\dif\upsilon$ the appropriate volume element.
We use in the following the above relation to estimate the errors on the FF parameters for one and two parameters expression of the FFs $\tilde f_+(t)$ and $\tilde f_0(t)$. The errors in any realistic experiment will be larger than our estimates, typically two to three times.  The above estimates are useful for  the understanding of the problems in the determination of the parameters in question.
The elements of G depend on the values of the parameters.
 In the case of the form factors, the errors on the $\lambda$ parameters
 change insignificantly for 10\% changes of the parameters.
 In other words the errors do not depend on the data
 being fitted and the correlations apply also to the systematic part
 of the errors.
\subsection{$K_{e3}$ decays}
For a quadratic FF, $\tilde f(t)=1+\lamp(t/m^2)+(\lampp/2)(t/m^2)^2$, the inverse of
the covariance matrix ${\bf G}_+^{-1}$, the covariance matrix ${\bf G}_+$ and the correlation matrix are:
$$N\,\pmatrix{5.937 & 13.867\cr\noalign{\vglue-4mm}\cr 13.867 & 36.2405\cr},\kern5mm
{1\over N}\,\pmatrix{1.258^2 & -0.606 \cr\noalign{\vglue-4mm}\cr -0.606 & 0.509^2},\kern5mm
\pmatrix{1 & -.945\cr\noalign{\vglue-4mm}\cr  & 1\cr}$$
The square root of the diagonal elements of ${\bf G}_+$ gives the errors, which for one million events are \dlp=0.00126, \dlpp=0.00051. The correlation is very close to \minus1, meaning that, because of statistical fluctuation of the bin counts, a fit will trade \lamp\ for \lampp\ and that the errors are enlarged.
A fit for a linear FF, $\tilde f(t)=1+\lamp(t/m^2)$ in fact gives \lamp=0.029 instead of 0.025 and an error smaller by \ab3:
$$\dlp=\sqrt{{\bf G}_+(1,1)}=0.0004.$$
A simple rule of thumb is that ignoring a $t^2$ term, increases \lamp\ by \ab3.5\x\lampp.
For $K_{e3}$ decays the presence of a $t^2$ term in the FF is firmly established. It is however not fully justified to fit for two parameters connected by the simple relation \lampp=2\x\lamp\up2. The authors of ref. \cite{stern} explicitly give an error for their estimate of the coefficient of the $t^2$ terms. 
\subsection{$K_{\mu3}$ decays}
The scalar FF only contributes to $K_{\mu3}$ decays. Dealing with these decays is much harder because: a) - the branching ratio is smaller, resulting in reduced statistics, b) - the $E_\pi$ or $t$ range in the decay is smaller, c) - it is in general harder to obtain an undistorted spectrum and d) - more parameters are necessary. This is quite well evidenced by the wide range of answers obtained by different experiments \cite{KTeV+04:FF,ISTRA+04:m3FF,NA48+06:m3FF,KLOE+07:m3FF}.
Assuming that both scalar and vector FF are given by quadratic polynomials as 
in Eq.~(\ref{eq:Taylor}), ordering the parameters as \lamop, \lamopp, \lamp\ and \lampp, the matrices ${\bf G}_{0\,\&\,+}^{-1}$ and ${\bf G}_{0\,\&\,+}$, are:
$$N\pmatrix{1.64&5.44&1.01&3.90\cr\noalign{\vglue-4mm}\cr
5.44&18.2&3.01&12.3\cr\noalign{\vglue-4mm}\cr
1.01&3.01&1.47&4.24\cr\noalign{\vglue-4mm}\cr
3.90&12.3&4.24&13.8\cr},\ {1\over N}\pmatrix{63.9^2&-1200&-923&197\cr\noalign{\vglue-4mm}\cr
-1200&18.8^2&272&-59\cr\noalign{\vglue-4mm}\cr
-923&272&14.8^2&-49\cr\noalign{\vglue-4mm}\cr
197&-59&-48&3.4^2 \cr}$$
and the correlations, ignoring the diagonal terms, are:
\begin{equation}
\label{4corr}
\pmatrix{-0.9996&-0.974&0.91\cr\noalign{\vglue-4mm}\cr
 &0.978&-0.919\cr\noalign{\vglue-4mm}\cr
 & &-0.976\cr}.
\end{equation}
All correlations are very close to \minus1. In particular the correlations between \lamop\ and \lamopp\ is \minus99.96\%, reflecting in vary large \dop\ and \dopp\ errors. We might ask what the error on \lamop\ and \lamopp\ might be if we had perfect knowledge of \lamp\ and \lampp. The inverse covariance matrix is give by the elements (1,1), (1,2), (2,1) and (2,2) of the ${\bf G}_{0\,\&\,+}^{-1}$ matrix above. The covariance matrix therefore is :
$${\bf G}_{0}(\lamop,\ \lamopp\ \hbox{for \lamp, \lampp\ known})={1\over N}\pmatrix{8.2^2&-20\cr\cr\noalign{\vglue -4mm}-20&2.4^2\cr}.$$
For one million events we have \dopp=0.0024, about 4\x\ the expected value of \lamopp. In other words \lamopp\ is likely to be never measurable. It is however a mistake to assume a scalar FF linear in $t$, because the coefficient of $t$ will absorb the coefficient of a $t^2$ term, again multiplied by \ab3.5. Thus a real value \lamop=0.014 is shifted by the fit to 0.017, having used the parametrization of Ref.~\cite{stern}.

\subsection{From the linear to the dispersive parametrization}
The results on the FFs obtained with the linear parametrization
can be used to determine the parameter of the dispersive
parametrization. As shown in the previous section the correlation
between \lamop\ and \lamopp\ is close to -1 and any fit to \lamop,\ \lamopp\
form  $K_{\mu3}$ decays will give values satisfying the relation:
\begin{equation}
  \lamopp = \tan\phi\lamop+B;
\end{equation}
$\tan\phi$ is independent on the number of events of the
 experiment
\begin{equation}
      \tan2\phi= \frac{2\rho_{1,2}\sigma_{2}\sigma_{1}}{\sigma^2_{2}-\sigma^2_{1}}~{\rm with}~  1= \lamop,~2=\lamopp\
\end{equation}
and amounts to -0.3. B can be determined from 
the experimental results for \lamo\ obtained using the linear
 parametrization ( \lamopp=0). 

\begin{figure}[htb]
\centering
\includegraphics[width=0.5\textwidth]{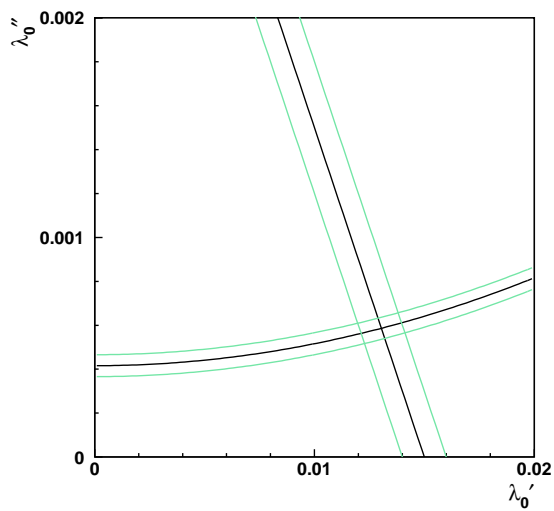}
\caption{\label{fig:trasla}Linear parametrization
 extrapolation along correlation line and relation from dispersive parametrization.
 }\end{figure}

 Therefore we can translate the results for \lamo\ in 
 any new parametrization with only one parameter
 with almost negligible 3th order term. In particular
 the dispersive parametrization gives:
\be
\lamopp = \lamop^2+(4.16\pm0.50)\times 10^{-4}
\ee
the procedure is shown in figure~\ref{fig:trasla}
for \lamo=$(15\pm 1)\times10^{-3}$.

}

\end{document}
